\documentclass[11pt,a4paper]{article}

\pdfoutput=1
\usepackage{jcappub}

\usepackage[english]{babel}  
\usepackage[T1]{fontenc}
\usepackage[load-configurations = abbreviations]{siunitx}
\usepackage{url}
\usepackage{xfrac}
\usepackage[dvipsnames]{xcolor}
\usepackage{float}
\usepackage{multirow}

\hypersetup{urlcolor=RedViolet,
	    citecolor=ForestGreen ,
	    linkcolor=RoyalBlue, 
	    colorlinks=true
}

\let\vec\mathbf %bold vector instead of arrow
\newcommand{\be}{\begin{equation}}
\newcommand{\ee}{\end{equation}}
\newcommand{\bea}{\begin{eqnarray}}
\newcommand{\eea}{\end{eqnarray}}

\newcommand{\kev}{\rm{keV}}
\newcommand{\gagg}{g_{a\gamma\gamma}}

\newcommand{\agg}{a\gamma\gamma}

\title{Solar axions in large extra dimensions}

\author[a]{Mar Bastero-Gil,}
\author[b]{Cyprien Beaufort,}
\author[b]{Daniel Santos}

\affiliation[a]{Departamento de Física Teórica y del Cosmos, Universidad de Granada, Granada-18071, Spain}
\affiliation[b]{Laboratoire de Physique Subatomique et de Cosmologie, Univ. Grenoble-Alpes(UGA), CNRS, Grenoble INP$^*$, LPSC-IN2P3, 38000 Grenoble, France}
\affiliation[*]{Institute of Engineering Univ. Grenoble Alpes}

\emailAdd{mbg@ugr.es}
\emailAdd{beaufort@lpsc.in2p3.fr}
\emailAdd{santos@lpsc.in2p3.fr}

\abstract{The axion could be used as a probe for extra dimensions. In large extra dimensions, besides the QCD axion one obtains an infinite tower of massive Kaluza-Klein (KK) states. We describe the processes of KK axions production in the Sun via the axion-photon coupling, $g_{a\gamma\gamma}$, and we derive the number density of KK axions that get trapped into the solar gravitational field and then accumulate over cosmic times. The large multiplicity of states, as well as their masses in the keV-range, deeply alter the phenomenology of the axion. This scenario leads us to propose the presence of KK axions as an interpretation of the non-thermal distribution of the solar X-rays. In this work, we dedicate special attention on the astrophysical and cosmological bounds that apply to the model. In particular, we show how the KK axions may escape the EBL limit that constrains standard ALPs in the same mass range. Present searches for KK axions make use of the decay channel, $a\rightarrow\gamma\gamma$, for which we revise the event rate; our value lies orders of magnitude below the rate usually quoted in the literature. This major conclusion stems from recent measurements of the luminosity of the quiet Sun which acts as an irreducible limit. The revised model remains a viable and an attractive explanation for multiple astrophysical observations, and we propose several approaches to search for solar KK axions in the near future.
}

\arxivnumber{}

\begin{document}

\maketitle

\section{Introduction}

In theories with large extra dimensions, the fundamental scale of quantum gravity can be lowered near the TeV-scale without violating experimental constraints \cite{ADD, ADDpheno}. Such a framework naturally solves the mass hierarchy problem and it can be embedded in string theory \cite{Witten1996, Lykken1996, Antoniadis1998} while preserving gauge coupling unification \cite{DienesUnification, Dumitru1999}. This approach offers new models describing, among others, inflation \cite{DvaliInflation, ADDinflation}, the baryon asymmetry \cite{DvaliBaryon} or the proton stability \cite{ADDstability}. In the original scenario, called ADD from the names of the founders, the Standard Model (SM) particles are constrained to live on a $4$-dimensional Minkowski brane whereas singlets under the SM gauge group can propagate into the bulk of dimension $(4+n)$. The weakness of gravity at large distances is explained by the propagation of gravitons in the bulk. The $n$ extra dimensions are compactified and large compared to the electroweak scale, up to the micrometer range, which allows for experimental tests of the theory.

The ADD's claim that the electroweak scale is the only fundamental scale seems, at first glance, incompatible with the usual need for intermediate scales to address phenomenological issues as the neutrino oscillations or the strong CP problem. However, the potential propagations of right-handed neutrinos and axions in the bulk give rise to a higher-dimensional seesaw mechanism without a heavy mass scale \cite{DienesNeutrino, ADDneutrino}, and to an experimentally allowed breaking of the Peccei-Quinn (PQ) symmetry near the electroweak scale that preserves axion invisibility \cite{Chang1999Long, Dienes, DiLella2000, ADDpheno}. 

In the standard picture, the axion is a hypothetical pseudoscalar that arises from the breaking of a global $U(1)$ symmetry called the PQ symmetry \cite{Peccei1977, Weinberg1978, Wilczek1978}. When the axion relaxes to its minimum of the potential, it dynamically compensates the CP-violating term of QCD and consequently solves the strong CP problem. The scale of the PQ symmetry breaking, $f_{PQ}$, is assumed to be in the range $f_{PQ}\sim(10^9 - 10^{12})\,\mathrm{GeV}$ which makes the axion a stable particle, weakly coupled to the SM particles and with a mass $m_{PQ}\sim \mathcal{O}(10\,\mathrm{\mu eV})$. The axion is hence a viable non-baryonic candidate for Dark Matter \cite{Preskill:1982cy,Abbott:1982af,Dine:1982ah} even if this interesting feature was not the initial purpose of its prediction.

The situation drastically changes in the framework of large extra dimensions. In the simplest scenario, the $n$ extra dimensions are toroidally compactified and we assume axions are free to propagate in $\delta \le n$ extra dimensions. Due to the periodicity of the extra dimensions, the axion field in the bulk can be Fourier expanded into an infinite number of modes. From our 4-dimensional brane, the axion is seen as an infinite superposition of massive states, called a Kaluza-Klein (KK) tower, and there is one KK tower for each of the $\delta$ extra dimensions in which the axion propagates. In other words, observed from our 4-dimensional brane, the standard QCD axion appears as complemented by $\delta$ infinite towers of massive KK states arising from the new phase space available in the bulk.

The presence of the KK towers and the extra dimensions deeply affect the physics of the axion. First, the coupling of the KK axions remains parametrized by $f_{PQ}$ whereas the overall scale of the PQ symmetry breaking, $\bar{f}_{PQ}$, is now suppressed by a volume-renormalization factor. While the experimentally constrained scale $f_{PQ}$ is unchanged, $\bar{f}_{PQ}$ can lie near the electroweak scale. Second, only the ground state of the KK tower, $a_0$, transforms under a PQ transformation \cite{Dienes}. The strong CP problem is still solved in extra large dimensions and $a_0$ can be identified as the standard QCD axion. Third, KK axions would be produced with masses in the keV-range. Each KK mode can decay into 2 photons with lifetimes that can be of the order of the age of the universe, which leads to important consequences on astrophysics and cosmology, but also on detection since the channel $a\rightarrow \gamma\gamma$ has been proposed for experimental searches \cite{DiLella2000, DiLella2003}. Finally, the large number of states strongly enhances any phenomenon mediated by the axions. 

KK axion modes up to the kinematic limit would be produced in astrophysical objects. The Sun appears as a particularly interesting source due to its proximity and its relatively well-known characteristics. Moreover, part of the KK axions produced by the Sun are sufficiently non-relativistic to be trapped into the solar gravitational field. These orbiting axions accumulate over cosmic times and consequently increase the present KK axion flux in a detector on Earth by about 5 orders of magnitude compared to the direct flux of axions emitted by the Sun. The decay of these trapped KK axions into photons acts as a new X-rays source in the solar system and has been proposed as an explanation to the coronal heating problem \cite{DiLella2003} and to the non-thermal distribution of the solar X-rays measurements \cite{Zioutas2004}.

Searches for solar KK axions have started using axion helioscopes \cite{Horvat2004, Lakic2008} or via the decay into photons \cite{Naples2004, Morgan2005, Battesti2007, XMASS, PacoThesis}. As far as we know, there is only a single paper describing the model of trapped KK axions \cite{DiLella2003} which sets the basis for experimental searches, and this pioneer paper contains few mathematical details. The present work aims to entirely revise the solar KK axion model, to provide analytical expressions for phenomenology, and to update the experimental constraints that apply on the model with recent measurements. We expect to shine light on how KK axions can be used as probe for extra dimensions. 

The paper is organized as follows. In Section \ref{sec:axionLED} we describe the physics of the axions in large extra dimensions. Section \ref{sec:axionProd} is dedicated to processes of KK axions production in the Sun in order to determine the solar KK axions flux. The case of gravitational trapped KK axions is discussed in Section \ref{sec:trappedAxions} where we derive an expression for the number density of KK axions, which plays a major role for experimental searches. In Section \ref{sec:constraints}, we revise the constraints that apply on the model and we discuss a possible explanation of the non-thermal distribution of the solar X-rays measurements. Finally, we conclude by studying some strategies of detection in Section \ref{sec:searches}. In particular, we update the decay event rate ($a\rightarrow\gamma\gamma$) in a detector on Earth showing that this channel of detection, the most investigated in experiments, seems inaccessible with present technologies, except if one can reject almost perfectly the background. We anyway propose other approaches to search for solar KK axions. 

%%%%%%%%%%%%%%%%%%%%%%%%%%%%%%%%%%%%%%%%%%%%%%%%%%%%%%%%%%

\section{Axions in large extra dimensions}\label{sec:axionLED}

Let first briefly describe the standard axions before discussing the higher-dimensional case. Axions are pseudo Goldstone bosons associated to the breaking of a global $U(1)_{PQ}$ symmetry, as proposed by Peccei and Quinn in order to solve the strong CP problem \cite{Peccei1996}. For the moment we limit our study to the minimal scenario of hadronic models in which axions do not couple to fermions at tree-level. The effective axion Lagrangian of interest for us is given by:

\begin{equation}
    \mathcal{L}_{eff} ~=~ \frac{1}{2}\big(\partial_\mu a\big)^2\, -\, \frac{1}{2}m_{PQ}^2a^2\,+\, \frac{g_{a\gamma\gamma}}{4}a\,F^{\mu\nu}\tilde{F}_{\mu\nu} \,,
    \label{eq:Lag4D}
\end{equation}
in which $a$ is the axion, $F^{\mu\nu}$ is the electromagnetic field tensor, $\tilde{F}_{\mu\nu}$ is its dual, and $m_{PQ}$ is the axion mass that can be expressed as \cite{diCortona2015}:
\begin{equation}
    m_{PQ} = 5.70(7)\, \mu\mathrm{eV}~\bigg(\frac{\num{e12}\,\mathrm{GeV}}{f_{PQ}}\bigg) \,. 
\end{equation}
We call $f_{PQ}$ the scale of the $U(1)_{PQ}$ symmetry breaking and we define the effective axion-photon coupling $g_{a\gamma\gamma}$ as:
\begin{equation}
    g_{a\gamma\gamma} \equiv \frac{\xi \alpha_{em}}{2\pi f_{PQ}}\,,
    \label{eq:gayyDef}
\end{equation}
where $\xi$ is a model-dependent constant of order unity and $\alpha_{em}$ is the fine-structure constant. The scale of symmetry breaking, $f_{PQ}$, is constrained from astrophysical (lower limit) \cite{Vysotsky:1978dc,Raffelt} and cosmological (upper limit) considerations \cite{Preskill:1982cy,Abbott:1982af,Dine:1982ah,Berezhiani:1992rk}, with the usually quoted bounds:
%r
\begin{equation}
	10^9\,\mathrm{GeV} \lesssim f_{PQ} \lesssim 10^{12}\,\mathrm{GeV} \,. 
	\label{eq:Raffelt}
\end{equation}
Finally, we present the expression for the lifetime of the axion:
\begin{equation}
    \tau_{a\rightarrow\gamma\gamma}~=~\frac{64\pi}{g_{a\gamma\gamma}^2m_{PQ}^3} ~\simeq~ 1.5\times10^{45}\,\mathrm{days}~\bigg(\frac{10^{-15}\,\mathrm{GeV^{-1}}}{g_{a\gamma\gamma}}\bigg)^2\,\bigg(\frac{10^{-5}\,\mathrm{eV}}{m_{PQ}}\bigg)^3 \,. 
\end{equation}
From the previous expressions we see that the axion is feebly coupled to Standard Model particles, that it is a light particle and that its lifetime in the allowed range given by Eq.(\ref{eq:Raffelt}) is much larger than the age of the universe, rendering the axion effectively stable. This consideration rules out the possibility of detecting the axion through the decay channel $a\rightarrow\gamma\gamma$.

We will now discuss the higher-dimensional case and the consequences for axion physics. Let consider $n$ large extra spatial dimensions to be compactified on a $\mathbb{Z}_2$ orbifold with an orbifold action $\vec{y}\rightarrow-\vec{y}$. For simplicity and conceptual reasons we consider all the extra dimensions to have the same compactification radius $R$. In this scenario, the graviton is free to propagate in all dimensions and the fundamental gravitational scale, $M_\ast$, can be much lower than the Planck scale, $M_P$, due to the extra dimensions volume suppression. They are related to each other via the expression \cite{ADD}:
\begin{equation}
	M_P = \big(2\pi R M_\ast\big)^{n/2}\,M_\ast \,. 
	\label{eq:volSupp}
\end{equation}
The original motivation of such a scenario was to keep the fundamental scale $M_\ast$ in the TeV range in order to solve the hierarchy problem of the SM. While the particles of the SM are constrained to live on a $4$-dimensional Minkowski brane, any singlets under the SM symmetry may propagate in $\delta \leq n$ dimensions. This can be the case for the axion and, similarly to the gravitational case, the $4$-dimensional axion scale $f_{PQ}$ can be much larger than the fundamental axion scale $\bar{f}_{PQ}$ of the overall theory:
\begin{equation}
	f_{PQ} = \big(2\pi R M_\ast\big)^{\delta/2}\,\bar{f}_{PQ}\,. 
	\label{eq:fPQBulk}
\end{equation}
This relation tells us that the fundamental scale $\bar{f}_{PQ}$ could be very low, perhaps in the TeV range, while letting the experimentally constrained 4-dimensional scale $f_{PQ}$ in the range of Eq.(\ref{eq:Raffelt}). It is also tempting to keep the fundamental scale $M_\ast \sim \mathcal{O}(1-100\,\mathrm{TeV})$ in order to solve the mass hierarchy problem of the Standard Model. Without introducing any other fundamental besides $M_\ast$ and the compactification radius $R$, \textit{i.e.} taking $\bar{f}_{PQ} = M_\ast$, we see that if the axion propagates in the same number of extra dimensions than gravity, the 4-dimensional axion scale $f_{PQ}$ must be at the Planck scale and thus disagree with the experimental constraints of Eq.(\ref{eq:Raffelt}).

The size of the extra dimensions, described by $R$, influences many observables and is consequently constrained from multiple measurements. A considerable review of such limits can be found in \cite{PDGLED}, we list in Table \ref{tab:R} the most striking bounds on $R$ obtained from tests of Newton law, colliders  and astrophysical considerations.

\begin{table}[H]
\hspace*{-1.5cm}
{\setlength{\extrarowheight}{4pt}%
\begin{tabular}{ccccccccc}
\hline \hline
\multicolumn{2}{c}{$n$}                                          & 1 & 2 & 3 & 4 & 5 & 6 \\ \hline
\multirow{4}{*}{$R_{max}\,[\mathrm{keV}^{-1}]$} 
& Pendulum & $2.2\times 10^5$ & / & / & / & / & /   \\
										& 
Colliders   & /	& $2.5\times 10^{4}$  & $6.2\times 10^{-1}$   & $3.1\times 10^{-3}$  &  $1.2\times 10^{-4}$ &  $1.5\times 10^{-5}$   \\
												&
SN 1987A & $2.5\times 10^{12}$ &$4.9\times 10^{3}$  & $5.8$   & $1.9\times 10^{-1}$  &  $2.5\times 10^{-2}$ &  $6.1\times 10^{-3}$ \\
                                                & Neutron star & $2.2\times 10^5$ &$7.9\times 10^{-1}$  & $1.3\times 10^{-2}$   & $1.7\times 10^{-3}$  &  $5.1\times 10^{-4}$ &  $2.2\times 10^{-4}$ \\ \hline \hline
\end{tabular}}
\caption{Limits on $R$ from torsion-pendulum experiments \cite{Kapner2007}, from colliders (CMS) \cite{Sirunyan2017} and from the duration of the neutrino signal of the supernova SN 1987A and neutron-star excess heat \cite{Hannestad2003}.}
\label{tab:R}
\end{table}

The results quoted as "neutron stars" set the most stringent limits on $R$ for $\delta<5$. They are obtained from the hypothetical $\gamma$ rays production due to the decay of KK gravitons trapped in a neutron star gravitational field. However, we suppose here the existence of KK axions, and it has been shown that KK gravitons would mainly decay into KK axions if they exist \cite{Chang1999Long}, process that would alter the derived limits. Plus, the branching ratio of KK graviton decay into photons could be drastically lowered if another brane exists as pointed out by the founders of the ADD model \cite{ADDpheno}. For those reasons, we do not consider the limits on $R$ derived from neutron stars in this work. 

With those elements in mind, and following the approach of \cite{ADDpheno, Chang1999Long, Dienes, DiLella2000}, let us have a look to the physics of the axion in large extra dimensions as seen from our 4-dimensional brane. Let denote $\vec{y}=(y_1, y_2, \dots, y_\delta)$ the coordinates of the axion in each extra dimensions $\delta \leq n$ it propagates so that the coordinates of the axion can be written $x^M \equiv (x^\mu, \vec{y})$ where $M$ is a spacetime index that runs over the $(4+\delta)$ dimensions. The $(4+\delta)$-action is given by:
\begin{equation}
    \mathcal{S}_{4+\delta} ~=~ \int d^4x\,d^\delta\vec{y}\,\bigg\lbrace \frac{1}{2} M_\ast^\delta (\partial_Ma)(\partial^Ma) ~+~ \frac{\xi\alpha_{em}}{8\pi\bar{f}_{PQ}}\,a\,F_{\mu\nu}\tilde{F}^{\mu\nu}\,\delta^{(\delta)}(\vec{y})\,\bigg\rbrace \,.
    \label{eq:action}
\end{equation}
We now need to compactify the extra dimensions in order to get the effective 4-dimensional theory. Since the axion field is compactified on a $\mathbb{Z}_2$ orbifold, one has:
\begin{equation}
    a(x^\mu, \vec{y}) = a(x^\mu, \vec{y} + 2\pi R)\hspace*{0.5cm} \mathrm{and}\hspace*{0.5cm} a(x^\mu, \vec{y}) = a(x^\mu, -\vec{y}) \,,
\end{equation}
and therefore one can Fourier expand the field without loss of generality:
\begin{equation}
    a(x^\mu, \vec{y}) = \sum_{\vec{n}=0}^{\infty}~a_{\vec{n}}(x^\mu)\,\cos\Big(\frac{\vec{n}\,\vec{y}}{R}\Big) \,,
\end{equation}
yielding an infinite superposition of modes; this procedure is called a Kaluza-Klein (KK) decomposition. In the equation, $a_{\vec{n}}(x^\mu) \in \mathbb{R}$ are the KK modes, $\vec{n}=(n_1, n_2, \dots, n_\delta)$ is a $\delta$-dimensional vector that labels the KK modes, and the sum runs over all extra dimensions $\sum^\infty_{\vec{n}=0} \equiv \sum^\infty_{n_1=0}\,\sum^\infty_{n_2=0}\dots\sum^\infty_{n_\delta=0}   $. The next step is to plug this KK decomposition into the $(4+\delta)$-action and to integrate over the $\delta$ extra dimensions to get the effective 4-dimensional Lagrangian of the theory that describes the axion as seen from our brane:
\begin{equation}
    \mathcal{L}_{eff}^{4D} ~=~ \frac{1}{2}\,\sum^\infty_{\vec{n} =0}\big(\partial_\mu a_{\vec{n}}\big)\big(\partial^\mu a_{\vec{n}}\big)
    ~-~ \frac{1}{2}\,m_{PQ}^2a_0^2 
    ~-~ \frac{1}{2}\,\sum^\infty_{\vec{n\neq0}}\,\frac{\vec{n}^2}{R^2}\,a_{\vec{n}}
    ~+~ \frac{\xi\alpha_{em}}{8\pi f_{PQ}}\,\Bigg(\sum^\infty_{\vec{n}=0}\,r_{\vec{n}}a_{\vec{n}}\Bigg)\,F^{\mu\nu}\tilde{F}_{\mu\nu} \,,
    \label{eq:Lfinal}
\end{equation}
with $r_0=1$ and $r_{\vec{n}\neq0} = \sqrt{2}$ which are rescaling coefficients to ensure that the KK modes $a_{\vec{n}}$ have canonically normalized kinetic-energy terms \cite{Dienes}. 

A few comments are needed at this stage. First, a single axion propagating in the bulk will be seen from our 4-dimensional brane as a sum of $\delta$ infinite superpositions of massive states with same quantum numbers, the overall superposition being called a Kaluza-Klein tower. The mass of each KK mode depends on the value of the momentum of the axion along the extra dimensions, this momentum being quantized due the compactification of the extra dimensions. The KK axion masses are given by \cite{Dienes}:
\begin{equation}
    m_{a_0} = \min \big(m_{PQ}\,,\, \frac{1}{2R}\big) \hspace*{0.5cm} \mathrm{and} \hspace*{0.5cm} m_{a_{\vec{n}}} \simeq \frac{|\vec{n}|}{R} = \frac{\sqrt{n_1^2 + n_2^2 + \dots + n_\delta^2}}{R} \,. 
\end{equation}
In the phenomenologically interesting situations that we will consider later we have $m_{PQ} \ll 1/R$, so the fundamental mode of the KK tower is equal to the 4-dimensional Peccei-Quinn mass, $m_{a_0} = m_{PQ}$. It means that one can let $m_{PQ} \sim \mathrm{10\,\mu eV}$ as in the standard case while having larger masses in the KK tower.  We also see that the mass splitting of the KK tower is $\sim 1/R$. The differential mode multiplicity can reach very large values and is given by:
\begin{equation}
    \frac{dN}{dm} =  \frac{2\pi^{\delta/2}}{\Gamma[\delta/2]}\,R^\delta\,m^{(\delta-1)} \,. 
\end{equation}
Second, it can be shown that a higher-dimensional Peccei-Quinn mechanism exists and that only the fundamental mode $a_0$ transforms under the $U(1)_{PQ}$ symmetry \cite{Dienes}. In other words, the strong CP problem is still solved when the axion propagates in the bulk and one can identify the lightest KK mode $a_0$ to the standard QCD axion. Third, the last term of the effective Lagrangian Eq.(\ref{eq:Lfinal}) tells us that each individual KK mode couples to the electromagnetic tensor with the same coupling than in the 4-dimensional case, $\frac{1}{4}g_{a\gamma\gamma}$ defined in Eq.(\ref{eq:gayyDef}), up to a factor $r_{\vec{n}}$ that we will ignore for the rest of the paper for simplicity. The lifetime of each individual KK mode is now given by:
\begin{equation}
    \tau\big(a_{\vec{n}}\rightarrow\gamma\gamma\big) ~\simeq~ \frac{64\pi}{g_{a\gamma\gamma}^2m_{a_{\vec{n}}}^3} ~\simeq~ 4.9\times10^{17}\,\mathrm{s}~\bigg(\frac{10^{-13}\,\mathrm{GeV^{-1}}}{g_{a\gamma\gamma}}\bigg)^2\,\bigg(\frac{30\,\mathrm{keV}}{m_{a_{\vec{n}}}}\bigg)^3 \,. 
\end{equation}

We can now touch the attractiveness of such a framework. The axion was first proposed to solve the strong CP problem and it has received considerable interest for more than four decades, both from theoretical and experimental aspects \cite{Kim2008, Graham2015}. In theories with large extra dimensions, the axion remains a compelling solution of the strong CP problem and a viable dark matter candidate, but its phenomenology changes. Instead of having one light particle, we must now consider an infinite number of massive states (the Kaluza-Klein axions), forming a continuum for large enough extra dimensions, with a large multiplicity of states. 

As in the standard case, KK axions would be produced in astrophysical objects. We will show that the Sun would produced KK axions with masses distributed in the $(1-30)\,\mathrm{keV}$ range. With such masses, the lifetime of KK axions could be of the order of the age of the universe and the photons obtained from its decay $a_{\vec{n}}\rightarrow\gamma\gamma$ would have impact on cosmology and astrophysics, and would open a new channel for experimental searches. Moreover, as we will demonstrate, part of the KK axions produced by the Sun would be trapped in its gravitational field.

%%%%%%%%%%%%%%%%%%%%%%%%%%%%%%%%%%%%%%%%%%%%%%%%%%

\section{Production of KK axions in the Sun}\label{sec:axionProd}    

In hadronic models, KK axions are produced in the Sun via three dominant processes. The Primakoff process $\gamma + Ze \rightarrow Ze + a$ converts a photon into an axion in the electrostatic field of a nucleus or an electron. The second process is the coalescence of two photons $\gamma\gamma\rightarrow a$, which is kinematically suppressed for small masses (as for the QCD axion) but turns to be significant for producing KK axions with masses in the keV-range. In the thermal environment of the Sun, photons acquire a thermal "mass" and a longitudinal component. This phenomenon opens up a third production process from the decay of a transverse plasmon $\gamma_T \rightarrow \gamma_L + a$. It is worth to mention at this stage that the phenomenology of solar KK axions is almost exclusively driven by the process of photon coalescence since it produces non-relativistic KK axions that are susceptible to be trapped in the gravitational field of the Sun, as we will see in Section \ref{sec:trappedAxions}.

\subsection{Primakoff process}\label{subsec:photonPrimakoff}

The Primakoff transition rate for KK axions produced in the Sun is derived in \cite{Raffelt:1985nk,DiLella2000}
\be
\Gamma_{\gamma\rightarrow a}^{\mathrm{Primakoff}} ~=~ 
\frac{g_{a\gamma\gamma}^2T\kappa^2}{32\pi^2}~\frac{|\vec k|}{\omega} \int d \Omega \frac{| \vec k \times \vec p |^2}{(\vec k- \vec p)^2 ((\vec k-\vec p)^2 + \kappa^2)} \,, \label{gammaprima}
\ee
where $T$ is the temperature at production, $\vec p$ is the  axion momentum, $\vec k$ the photon one, and $\omega$ the photon energy; $\kappa$ is the Debye-Hückel screening scale, given in general by $\kappa^2= (4 \pi \alpha/T) \sum_j Z_j^2 n_j$, where $n_j$ is the number density of charged particles.
For small momenta, it can be expressed as:
\begin{equation}
	\Gamma_{\gamma\rightarrow a}^{\mathrm{Primakoff}} ~=~ \frac{g_{a\gamma\gamma}^2T\kappa^2}{32\pi}\bigg(\frac{8p^2}{3(\kappa^2 + m^2)} + \mathcal{O}(p^4)\bigg) \,,
\end{equation}
where $p=|\vec p |$ and $m$ is the axion mass. The differential axion flux at Earth, integrating over a standard solar model, is given by
\be
\Phi^{\mathrm{Primakoff}}_a ~=~ \frac{1}{ 4 \pi d_{ST}^2} \int_{\rm Sun } d^3\vec r
\Gamma_{\gamma\rightarrow a}^{\mathrm{Primakoff}} ~ \frac{E^2}{\pi^2} ~f_a^{eq}(E) \,,
\label{FluxPrimakoff}
\ee
where $d_{ST}$ is the Earth-Sun distance, $E$ the axion energy, and $f_a^{eq}(E)=(e^{E/T}-1)^{-1}$ is the Bose-Einstein axion distribution function. The authors of \cite{DiLella2000} arrived at the expression for the differential axion flux at Earth:
\begin{equation}
	\Phi^{\mathrm{Primakoff}}_a ~=~ \Big(4.20\times 10^{10}~\mathrm{cm^{-2}\,s^{-1}\,keV^{-1}}\Big)~ g_{10}^2\,\frac{E\,p^2}{e^{E/1.1} - 0.7}\big(1+0.02m\big) \,,
\end{equation} 
where $g_{10} = g_{a\gamma\gamma} \times 10^{10}\,\mathrm{GeV^{2}}$ is the dimensionless coupling to photon, and  $E$, $p$, and $m$ are given in keV.

\subsection{Photon coalescence}\label{subsec:photonCoal}

Let us review the production of axions from the photon coalescence process $\gamma \gamma \rightarrow a$ in a  thermal environment as such of the Sun. We start with the Boltzmann equation for the axion distribution function, $f_a$, including only production from the inverse decay \cite{Cadamuro2011}:
\bea
\frac{d f_a(E)}{dt} &=&\frac{1}{2E} \int \frac{d^3 k_1}{(2 \pi)^3 2 \omega_1}  \frac{d^3 k_2}{(2 \pi)^3 2 \omega_2} (2 \pi)^4 \delta^4(k_1 + k_2 - p) |M_\gamma|^2 f_1^{eq}(\omega_1) f_2^{eq}(\omega_2) ( 1 + f_a(E))    \\
&\simeq & f_a^{eq}(E) ~\frac{1}{2E} \int \frac{d^3 k_1}{(2 \pi)^3 2 \omega_1}  \frac{d^3 k_2}{(2 \pi)^3 2 \omega_2} (2 \pi)^4 \delta^4(k_1 + k_2 - p) |M_\gamma|^2 (1+ f_1^{eq}(\omega_1) +f_2^{eq}(\omega_2)) \\
&=& \Gamma_{\gamma\gamma\rightarrow a}^{\rm{Coal.}}(T) f_a^{eq}(E) \,,
\eea
where $E$ is the axion energy, $p$ its momentum, and $\omega_i$ are the photon energies, with $E=\omega_1 + \omega_2$. In the the second line we have used:
\be
f_1^{eq}(\omega_1) f_2^{eq}(\omega_2) (1 + f_a^{eq}(E)) = f_a^{eq}(E)\cdot ( 1 + f_1^{eq}(\omega_1) +f_2^{eq}(\omega_2) ) \,,
\ee
for the Bose-Einstein distribution functions. Performing the momentum integrals, the thermal inverse decay rate, $\Gamma_{\gamma\gamma\rightarrow a}^{\rm{Coal.}}(T)$, is given by \cite{Cadamuro2011, Cadamuro2012}:
\be
\Gamma_{\gamma\gamma\rightarrow a}^{\rm{Coal.}}(T) = \Gamma_{\agg} \frac{m^2 - 4 \omega_P^2}{m^2} \left(\frac{m}{E}\right) \left( 1 + \frac{2T}{p} \ln \frac{1- e^{-(E+p)/2T}}{1- e^{-(E-p)/2T}} \right) \label{eq:GammaCoal} \,,
\ee
where $\Gamma_{\agg}$ is the standard axion decay rate into photons at $T=0$, and  $\omega_P^2$ the photon thermal "mass" that can be expressed as:
\be
\omega_P^2 = \frac{4 \pi \alpha n_e}{m_e} \,,
\ee
$n_e$ being the electron number density and $m_e$ its mass. Finally, for the number of axions per unit volume, per unit time, and per unit energy, we make use of $dp/dE = E/p$ to obtain:
\be
\frac{d N_a^{\rm{Coal.}}}{d E} = \Gamma_{\gamma\gamma\rightarrow a}^{\rm{Coal.}}(T)  \,\frac{p E}{2 \pi^2}\, f_a^{eq}(E) \,.
\ee
We will apply approximations for later analytical calculations in order to simplify the expressions, as done in \cite{DiLella2000}. Neglecting the photon thermal "mass" and the temperature corrections, Eq.(\ref{eq:GammaCoal}) simplifies into:
\be
\Gamma_{\gamma\gamma\rightarrow a}^{\rm{Coal.}} ~\simeq ~ \Gamma_{\agg}\frac{m}{E}\label{eq:coalRateSimplify}\,. 
\ee
Integrating over a standard solar model (in this work we use the Saclay solar model \cite{SaclayModel}), we obtain the differential flux at Earth for axions produced by the coalescence of two photons:
\bea
\Phi^{\rm{Coal.}}_a &\simeq & \frac{1}{ 4 \pi d_{ST}^2} \int d^3r \frac{d N^{\rm{Coal.}}_a}{d E} ~=~ \frac{R_\odot^3}{ 4 \pi d_{ST}^2} 4 \pi \Gamma_{\agg} \frac{p m}{2 \pi^2} \int_0^1 d \bar r \bar r^2 f_a^{eq}(E) \\
&=& \Big(7.45\times 10^{11}~\mathrm{cm^{-2}\,s^{-1}\,keV^{-1}}\Big)~ g_{10}^2\,m^4\,p\int_0^1 d \bar r \bar r^2 f_a^{eq}(E)  \label{FluxCoales}\,,
\eea
where $g_{10}$ is the dimensionless coupling described before, and $m$ and $p$ are given in keV. Still, when performing the integral, this expression slightly differs from the one derived by Di Lella \textit{et al.} \cite{DiLella2000}
\be
\Phi_a \simeq  \Big(1.68\times 10^{9}~\mathrm{cm^{-2}\,s^{-1}\,keV^{-1}}\Big)~ g_{10}^2\,m^4 p\, \bigg( {\frac{10}{0.2+E^2}+1+6\times 10^{-4} E^3}\bigg) e^{-E} \,. \label{FluxCoalesDiLella}
\ee

\begin{figure}[H]
	\centering
	\includegraphics[width=0.8\linewidth]{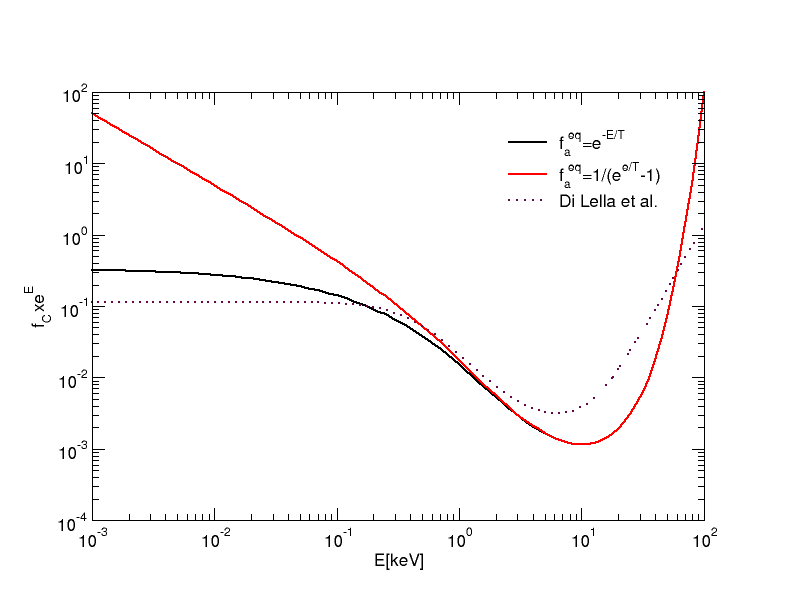}
	\caption{Energy dependency of the coalescence rate. We plot $f_C(E)\,e^E$ where $f_C(E) = \int_0^1 d \bar r \bar r^2 f_a^{eq}(E)$. The solid black line is obtained with a Maxwell-Boltzmann distribution, and the solid red line with a Bose-Einstein distribution. The dotted purple line shows the approximation derived in \cite{DiLella2000}.}
	\label{fig:coalescenceComparison}
\end{figure}

For comparison, Fig. \ref{fig:coalescenceComparison} shows the energy dependency of the coalescence rate in Eq. \eqref{FluxCoales} and \eqref{FluxCoalesDiLella}. For example,  in the energy range of interest, $E\sim \mathcal{O}(1-10)\,\mathrm{keV}$, the approximations applied by Di Lella \textit{et al.} would overestimate the flux of photons by about 25\% after integrating in the case $\delta = 1$. 

\subsection{Plasmon decay}\label{subsec:plasmonDecay}

In the thermal environment of the Sun, the excitations of the electromagnetic field follow non-trivial dispersion relations. These excitations, either longitudinal or transverse, are called "plasmons". The presence of plasmons in the solar interior opens up the possibility of producing KK axions by decay of a transverse plasmon, $\gamma_T  \rightarrow \gamma_L + a$, since this process becomes kinematically allowed. In a thermal plasma, transverse and longitudinal photons have different dispersion relations:
\bea
\omega_T^2 &=& \omega_P^2 \Big( 1 + \frac{k_T^2}{\omega_P^2 + k_T^2} \frac{T}{m_e}\Big) + k_T^2 ~\simeq~ k_T^2 + \omega_P^2 \,, \label{omegaT}\\
\omega_L^2 &=& \omega_P^2 \Big( 1 + 3\frac{k_L^2}{\omega_P^2} \frac{T}{m_e}\Big) ~\simeq~ \omega_P^2  \label{omegaL}\,,
\eea
where $k_{T,\,L}$ are the transverse and longitudinal 3-momenta, and $\omega_P$ is the plasma frequency defined previously. In the Sun, $T \lesssim 1.3$ keV  and  $T/m_e \lesssim 0.0025$, so the dispersions relations can be well approximated by the second equality in Eqs. \eqref{omegaT} and \eqref{omegaL}.

We will compute the rate of production of axions with momentum $p$ and energy $E$ from:
\be
\frac{d N_a^{\rm{Decay}}}{d E} = \Gamma_{\gamma_T\rightarrow a\gamma_L}^{\rm{Decay}}(T)  \frac{p E}{2 \pi^2} f_a^{eq}(E) \,,
\ee
where 
\be
\Gamma_{\gamma_T\rightarrow a\gamma_L}^{\rm{Decay}} (T) = \frac{1}{2E} \int \frac{d^3 k_T}{(2 \pi)^3 2 \omega_T}  \frac{d^3 k_L}{(2 \pi)^3 2 \omega_L} (2 \pi)^4 \delta^4(k_T - k_L - p) |M_{\gamma_T\rightarrow a\gamma_L}|^2 (1+ f^{eq}(\omega_T)) \label{eq:thermaldecTL}\,,
\ee
and the transition amplitude is given by \cite{RaffeltPlasmon}: 
\be
|M_{\gamma_T\rightarrow a\gamma_L}|^2= g^2_{\agg}\frac{|(\vec e_T \times \vec k_T)\cdot \vec k_L|^2}{k_L^2} \frac{\omega_P^4}{k_L^2 + \omega_P^2}
\,, 
\ee 
in which $\vec e_T$ is a polarization vector such that $\vec{e}_T \cdot\vec{k}_T = 0$. 

The details of the derivation of the plasmon decay rate are presented in Appendix \ref{app:plasmonDecay}. As it will be discussed in section \ref{sec:trappedAxions}, we are interested in non-relativistic KK axions so we restrict our analysis to the limit $p/k_T \ll 1$, for which we get:
\be
\Gamma_{\gamma_T\rightarrow a\gamma_L}^{\rm{Decay}} \simeq \frac{g^2_{\agg}}{24 \pi} \, \frac{\omega_P^4\,  k_T}{k_T^2 + \omega_P^2} \,\Big(\frac{p}{E}\Big)^2 \,. 
\ee
We now have all the elements to give the expression for the differential flux at Earth for KK non-relativistic axions produced by plasmon decay:
\be
\Phi^{\rm{Decay}}_a \simeq \Big(2.0\times 10^{12}~\mathrm{cm^{-2}\,s^{-1}\,keV^{-1}}\Big)~\ g_{10}^2 \,\int_0^1 d \bar r \bar r^2\,\frac{\omega_P^4\,  k_T}{k_T^2 + \omega_P^2}\,\frac{p^3}{E} \, f_a^{eq}(E) \,.
\ee

\begin{figure}[t]
	\centering 
	\includegraphics[width=0.8\linewidth]{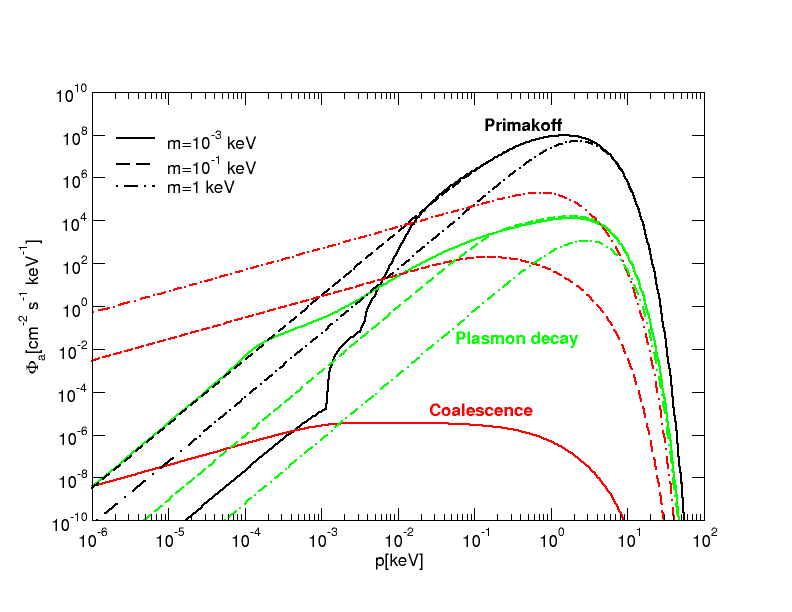}
	\caption{Differential flux at Earth for individual axions versus the axion momentum, for different production channels: black lines for Primakoff, red lines for Coalescence, and green ones for Plasmon decay. We have taken  $g_{10}=10^{-2}$, and different values of the  axion mass as indicated in the plot.}
	\label{fig:FluxesAll}
\end{figure}

The differential fluxes at Earth for Primakoff, Coalescence and Plasmon decay are shown in Fig. \ref{fig:FluxesAll}, for different values of the axion mass as indicated in the plot. The Primakoff differential rate has been computed with the transition rate in Eq. \eqref{gammaprima}, including the thermal photon mass  and using the approximation $\kappa \simeq 7 T$ \cite{DiLella2000}  for the Debye-Hückel screening scale; for Coalescence we use Eq. \eqref{FluxCoales}, and that of plasmon decay has been computed with the full thermal decay rate given in Eq. \eqref{EGammaTL}. We have plotted the fluxes with respect to the axion momentum instead of energy to stress their behavior at low momentum. Primakoff is the main production channel for axions upto $m\sim O(10)$ keV and $p \sim E$. However, it will be suppressed with respect to Coalescence in the limit of small momentum $p \ll E$. Same happens with the plasmon decay channel: in both cases the rate is proportional to $p^2$ which suppresses the production  of non-relativistic KK axions. 

We could also consider the production due to the conversion of  thermal photons in the presence of the large-scale solar magnetic field \cite{Mikheev:2009zz,Caputo:2020quz,Guarini:2020hps}. However, given the profile of the solar magnetic field, the conversion is mainly efficient near the resonance for $m \sim O(10-100) $ eV \cite{Guarini:2020hps}. Because we are interested in keV axions, this channel for us will be subdominant when compared to coalescence or Primakoff. 

\section{Gravitationally trapped KK axions }\label{sec:trappedAxions}

Part of the KK axions produced in the Sun will be sufficiently massive and non-relativistic to be trapped in the solar gravitational field. Those KK axions will start orbiting the Sun and they will accumulate over the Sun history. The large astrophysical times involved in this picture would result in a significant amount of trapped KK axions which overpasses by several orders of magnitude the amount of KK axions instantaneously emitted by the Sun. This scenario of trapped KK axions was first proposed by DiLella and Zioutas \cite{DiLella2003} to explain some astrophysical issues as the heating of the solar corona, but also as a window for a possible detection of KK axions in a Time Projection Chamber (TPC). Recently, the phenomenological implications of such "stellar basins" of massive particles has been studied in more details yielding to striking consequences on direct and indirect detection \cite{VanTilburg}.

The article from DiLella and Zioutas about trapped KK axions \cite{DiLella2003} pioneered the domain and its results (obtained by simulations for a given set of parameters) have been used as reference for all searches for KK axions \cite{Morgan2005, XMASS, PacoThesis}. This article is almost 20 years old and, as far as we know, no additional theoretical work has been conducted about trapped KK axions. This paper aims to cover this issue by deriving analytically some phenomenological quantities and by updating the astrophysical constraints on the model with recent measurements. 

   \subsection{Number density of KK axions}
   
   The number density of trapped KK axions as a function of the distance to the Sun represents a crucial quantity for the determination of the event rate in a detector and also for some astrophysical constraints as presented later in Section \ref{sec:constraints}. Since the Primakoff process and the plasmon decay are suppressed for low-momenta, it can be shown \cite{DiLella2003} that the amount of trapped KK axions is about 3 orders of magnitude larger for the production via the coalescence of two photons than for the other processes. For this reason, we only consider the coalescence of two photons in the following calculations. 
      
   The details of the calculations of the number density are presented in Appendix \ref{app:density}. The derivation starts by looking at the gravitational potential inside the Sun to determine the trajectories of the trapped KK axions. For instance, we show that in order to reach the Earth, the kinetic energy per unit mass of particles produced at Sun surface must be in the range $[0.995363, 0.995374]$ in units of $GM/R_\odot$. The next step is to integrate the Boltzmann equation over momenta for which the trajectories are bounded at a distance $r$ to the Sun. Finally, we integrate over the KK tower and time, taking into account the potential decay of the axions. We arrive to the following expression for the number density of trapped KK axions:
\bea
n_{KK}^{(T)}(t, r) &=&\Big(2.2 \times 10^{14} {\rm cm}^{-3} \Big)  \, g_{10}^2\, \frac{2 \pi^{\delta/2}}{\Gamma[\delta/2]} \int dm \,(R m)^\delta m^5\, \frac{1 - e^{-t_\odot \Gamma_{\agg}}} {t_\odot \Gamma_{\agg}} \nonumber\\
&& \times\, \int_0^1 \bar r_0^2 d \bar r_0\, \frac{1}{e^{E/T} - 1}\, \frac{m}{E}\, I_v[\bar r, \bar r_0] \,,
\eea
where $g_{10}=g_{\agg}\times 10^{10}\,\mathrm{GeV^{-1}}$, $\bar{r}$ is the distance to the Sun normalized by the solar radius, $m$, $R^{-1}$, $T$ and $E$ are given in keV, and the product $t_\odot \Gamma_{\agg}$ must be dimensionless; the second integral must be performed over a solar model. Finally, $I_v[\bar r, \bar r_0]$ is the integral over velocities (in the plane of the trapped orbit) of the probability density for an axion of energy $E$ to be at radius $\bar r$. In Appendix \ref{app:density} we give two expressions for the quantity $I_v[\bar r, \bar r_0]$ depending on which expression we use for the probability density: 
\bea 
I_v^0(\bar r_0, \bar r) & = & \frac{1}{2 \pi \bar r^4} \sqrt{2 \bigg(\bar \Phi_G(\bar r_0) - \frac{1}{\bar{r}}\bigg) } \,, \label{eq:Ivdelta}\\
I_v^1(\bar r_0, \bar r) &=& \int_{\bar v_{min}}^{\bar v_{max}} d \bar v \bar v^2 \frac{1}{10 \pi^2} \frac{\big(2 \bar \Phi_G(\bar r_0) - \bar v^2\big)^{7/2}}{\Big(\bar v^2 - 2 \big(\bar \Phi_G(\bar r_0) - \frac{1}{\bar{r}}\big)\Big)^{1/2}} \label{eq:IvP}\,,
\eea 
in which $\bar \Phi_G(\bar r_0)$ is the gravitational potential normalized by its value at Sun surface, and $\bar v$ is the velocity normalized by $(GM/R_\odot)^{1/2}$.

\begin{figure}[t]
	\centering
	\includegraphics[width=0.8\linewidth]{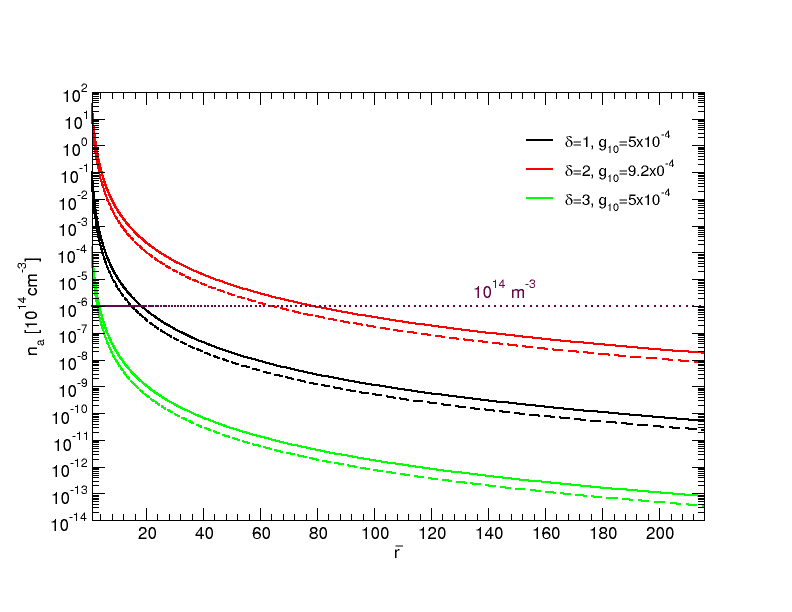}
	\caption{Number density of trapped KK axions for model parameters as indicated in the plot versus the radial distance in units of the solar radius. We have taken $R_1= 2 \times 10^5$, $R_2 =10^3$, $R_3=1$, all given in  keV$^{-1}$. Solid lines are obtained with \eqref{eq:Ivdelta}, while dashed lines correspond to \eqref{eq:IvP}. The dotted line indicates the value obtained in \cite{DiLella2003} for $n_{KK}^{(T)}(t_\odot,1 \,\rm{A.U.})$ and  $\delta=2$, $g_{10}=9.2\times10^{-4}$.}
	\label{fig:numberDensity}
\end{figure}

The number density behaves as $1/r^4$ as pointed out in \cite{DiLella2003}. For $\delta = 2$, $g_{10} = 9.2\times 10^{-4}$ and $R = 10^3\,\mathrm{keV^{-1}}$ (set of parameters used in \cite{DiLella2003}) we obtain
\bea
n_{KK}^{(T)}(t_\odot, R_\odot) &\simeq& 0.15 \times 10^{16} \, {\rm cm}^{-3}  \,,\\
n_{KK}^{(T)}(t_\odot,1 \,\rm{A.U.}) &\simeq& 0.79 \times 10^{6} \, {\rm cm}^{-3} \,, 
\eea
which is about 50 times lower than the number density at Earth obtained in \cite{DiLella2003}. Possible explanations of such a difference are difficult to point out since very few details are given in the article about how the simulation works. In Figure \ref{fig:numberDensity} we show the trapped number density for different choices of parameters, and we have compared the number density of trapped KK axions computed with the different probability densities (solid and dashed lines): there is only a factor of 2 difference. 
 
   \subsection{Monte Carlo simulation}
        
Since the number density plays a crucial role for phenomenology, we aim to cross-check the analytic calculations with a Monte Carlo (MC) simulation, both approaches using a different philosophy.

The backbone of the simulation is to produce a given amount of KK axions in the Sun and to follow the trajectory of each of them by solving the Equations of Motion (EoM). In this way, one knows the position of each KK axion at any time and one can compute the number density at a distance $r$. The number of KK axions of mass $m$ produced by the Sun evolves in time as:
\begin{equation}
	\frac{dN_a(t,m)}{dt} = -\Gamma_{a\gamma\gamma}(m)\,N_a(t,m) + P_a(m) \,, 
\end{equation}
where $P_a(m)$ is the solar production rate. It implies that, at present time $t=t_\odot$ and summing over the mode multiplicity, one obtains:
\begin{equation}
	N_a\big(\delta, R, g_{10}, t_\odot\big) = \frac{2\pi^{\delta/2}}{\Gamma[\delta/2]}\,R^{\delta}\int_0^\infty dm \, m^{(\delta-1)}\,  \frac{P_a(m)}{\Gamma_{a\gamma\gamma}(m)}\,\bigg(1-e^{-t_\odot\Gamma_{a\gamma\gamma}(m)}\bigg) \,.
\end{equation}

We have already seen that the coalescence mechanism is the main source of trapped axions. The solar axion production, per energy and time unit, through the coalescence process is given by:
\begin{equation}
	P_a(m, E) = \frac{g_{10}^2}{32\pi^2}\,m^4\,\sqrt{E^2 - m^2}\,\int_{Sun}\,dr\, \frac{r^2}{e^{E/T(r)} -1} \,,
	\label{eq:prodMC}
\end{equation}
in which the integration is over a solar model. In this work we use the Saclay solar model \cite{SaclayModel}. The present total number of KK axions susceptible of being trapped is then given by
\begin{equation}
\begin{split}
	N_a\big(\delta, R, g_{10}, t_\odot\big) = \frac{2\pi^{\delta/2}}{\Gamma[\delta/2]}\,&R^{\delta}\int_0^\infty dm \,  \frac{m^{(\delta-1)}}{\Gamma_{a\gamma\gamma}(m)}\,\bigg(1-e^{-t_\odot\Gamma_{a\gamma\gamma}(m)}\bigg) \\
	& \times \int_m^\infty \, dE\, \frac{g_{10}^2}{32\pi^2}\,m^4\,\sqrt{E^2 - m^2}\,\int_{Sun}\,dr\, \frac{r^2}{e^{E/T(r)} -1} \,.
\end{split}
\label{eq:NaMC}
\end{equation}
The principle of the MC simulation is to generate $n_{MC}$ KK axions of mass $m$ and energy $E$, to follow them in their trajectories around the Sun, and to count the proportion that will be located in a $1\,\mathrm{cm^3}$ box at distance $r$ to the Sun. Since the solar axion production is isotropic this box can be located at any position in the $(\phi, \theta)$ plane. We then scale $n_{MC}$ according to Eq.(\ref{eq:NaMC}) to get the present number density of KK axions.

The randomness of the MC comes from the choice of the initial conditions in the EoM solver: $r_0$ is taken randomly in the distribution of Eq.(\ref{eq:prodMC}); $\theta_0$ and $\phi_0$ are randomly chosen in a way to construct an isotropic angular distribution; and the duration of the integration, $t_{rand}$, is taken randomly in a power law distribution such that $t_{rand}$ is large compared to the orbit period.

\begin{figure}[H]
    \centering
    \includegraphics[width=0.9\linewidth]{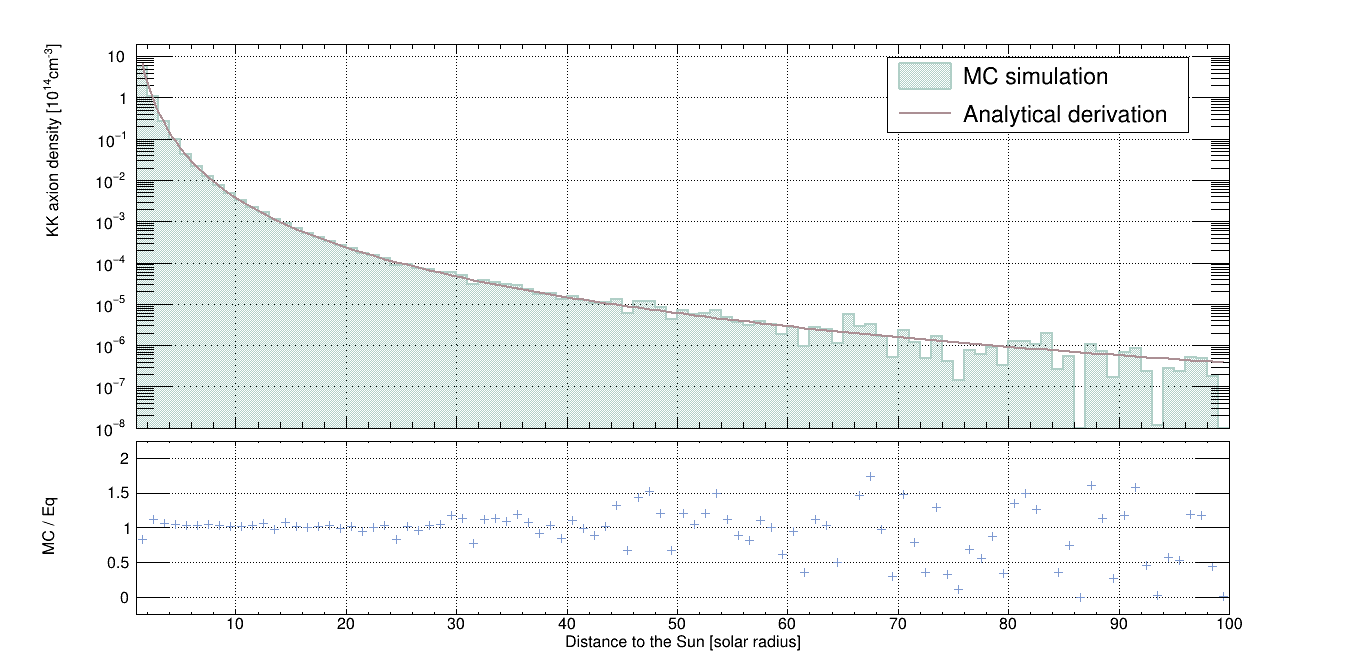}
    \caption{Comparison of the numbers density obtained from the MC simulation and from the analytical derivation. We have taken $\delta = 2$, $R = 10^3\,\mathrm{keV^{-1}}$ and $g_{10} = 9.2\times 10^{-4}$, which are the parameters cited in \cite{DiLella2003}.}
    \label{fig:compEqMC}
\end{figure}

A comparison between the MC simulation and the analytical derivation is presented in Figure \ref{fig:compEqMC}. For $r<40\,R_\odot$ the two approaches differ by less than 20\%. For larger $r$ the statistics of the MC becomes too low and we observe significant statistical fluctuations. This 20\% difference is small and is likely due to numerical uncertainties and approximations made in both approaches. The MC simulation agree on the order of magnitude obtained by the analytical derivation and thus it cross-checks the calculations.

The code of the MC simualtions is provided on a public repository \footnote{\url{https://github.com/BeaufortC/KKaxionDensity}}.

%%%%%%%%%%%%%%%%%%%%%%%%%%%%%%%%%%%%%%%%%%%%%%%%

\section{Constraining the solar KK axion model}\label{sec:constraints}

The existence of extra dimensions and of KK axions would alter some astrophysical and cosmological considerations. To be viable, the solar KK axions model must be in agreement with experimental constraints. In this section we review the constraints applying on the model in order to determine the parameter space in which KK axions should be searched for.

\subsection{Solar luminosity from KK axion}
    
The first constraint that we will consider is the one that puts the most stringent limits on the solar KK axion model. Paradoxically, it is also the one that brings the most interest in searching for KK axions to possibly explain the non-thermal X-ray emission of the Sun. As explained in Section \ref{sec:trappedAxions}, part of the KK axions gets trapped in the solar gravitational field and accumulates. A KK axion eventually decays into two photons and contributes to the solar luminosity. This solar luminosity due to trapped KK axions should not exceed the observed solar luminosity, requirement that sets constraints on the parameters of the model.

The maximal value for the axion-photon coupling obtained by DiLella and Zioutas using ASCA/SIS X-ray data \cite{DiLella2003} has been reduced by more than one order of magnitude few years later using measurements of the quiet Sun from RHESSI \cite{Hannah2007, Hannah2010}. While the solar axion luminosity is stable during short time periods, the standard solar luminosity depends on the Sun's activity. For this reason, we must compare our predictions to measurements of quiet Sun when the solar X-ray emissions are not related to flares, sunspots or active regions. In this work, we refer to solar X-rays measurements in the range $(1-6)\,\mathrm{keV}$ from the SphinX spectrophotometer \cite{Sylwester2012, Sylwester2019} during a period of deep solar minimum in 2009 known for its extremely low solar activity.
    
The SphinX spectrometer is orbiting the Earth and it has a conical field of view (FoV) of $2^\circ$. We want to determine the flux of photons coming from KK axion decays that enters the FoV of SphinX. Considering for simplicity that the detector covers a cone of aperture $\alpha_0$, we should integrate the number density of trapped KK axions over the solid angle of the detector, $d \Omega_D$, and over a radial distance $D$ from the detector (\textit{i.e.}, the Earth). Here we follow the approach in references \cite{Boyarsky_2006, Boyarsky_2007, Boyarsky_2008, Boyarsky_2008_SPI}. In these references, the flux of photons due to decaying dark matter (number of photons per unit time and unit surface) is given by:
\be
F_{DM} = \Gamma_{DM} \frac{E_\gamma}{m_{DM}} \int_{FoV} \frac{\rho_{DM} (r)}{4 \pi | {\bf r_D} + {\bf r}|^2} d^3 {\bf r} \,, \label{FDMFoV}
\ee
where $\Gamma_{DM}$ is the decay rate into photons, $r_D$ is the distance between the Sun and the Earth. The integral is done within the FoV of the detector. For example, for a small FoV, and DM decaying into two photons with $E_\gamma = m_{DM}/2$, one has: 
\be
F_{DM}[\alpha] \simeq \Gamma_{DM} \frac{\Omega_{fov}}{8 \pi} \int \rho_{DM} (r) dz \,,
\ee
where the integral is the column integral along the line of sight $z$, $r= r_D^2 + z^2- 2 r_D z \cos\alpha$, and $\alpha$ is the angle between $D$ and $r_D$ centered on the Earth.

In our case, the number of trapped KK axions within the detector FoV is given by (see Appendix \ref{app:FoV} for the details of the derivation):
\bea
N_a^{T,\,FoV}(m) &=& 
\int_{FoV}d {\bf r}^3 \frac{r^2}{4 \pi D^2} n_a^T(r,m) \, \nonumber \\
&\simeq & \frac{R_\odot n_a^T(R_\odot,m)}{2 \bar r_D^3}  \left( 2 \bar r_D - \frac{2}{3 \bar r_D} - \frac{1}{2}(1 - \frac{\alpha_0-\pi}{\tan \alpha_0} ) 
\right) \,, \label{NaTEarth}
\eea
and the flux of photons due to the decay of trapped KK axions and that enters the FoV of the detector by:
\be
\frac{d F}{ d E_\gamma} = 2 \frac{d F}{ d m} =2
\frac{2 \pi^{\delta/2}}{\Gamma[\delta/2]} (R m)^\delta  m^{-1} \Gamma_{\agg} N_a^{T,\,FoV}(m) \,. 
\ee

\begin{figure}[t]
	\centering
\begin{tabular}{c}
        %	\begin{minipage}{0.49\linewidth}
	\includegraphics[width=0.8\linewidth]{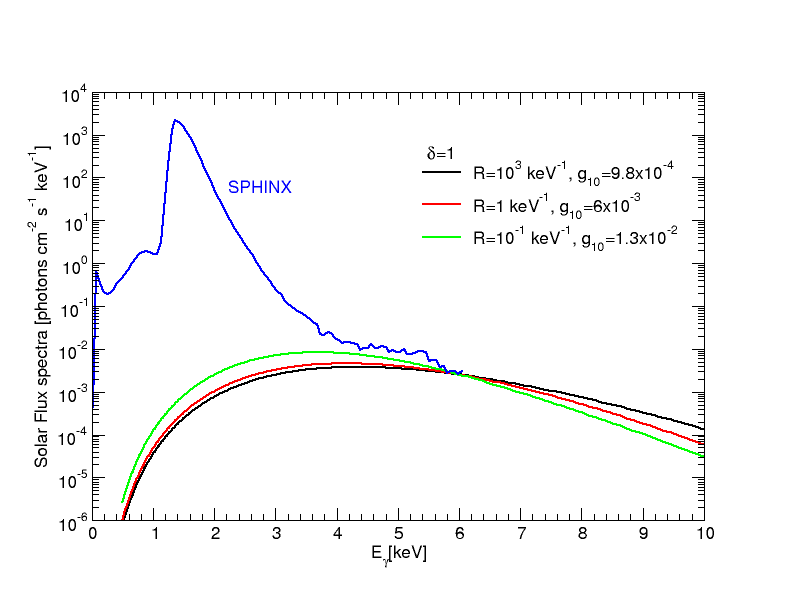} \\
%	\end{minipage}	
%	\hfill	
%	\begin{minipage}{0.5\linewidth}
	\includegraphics[width=0.8\linewidth]{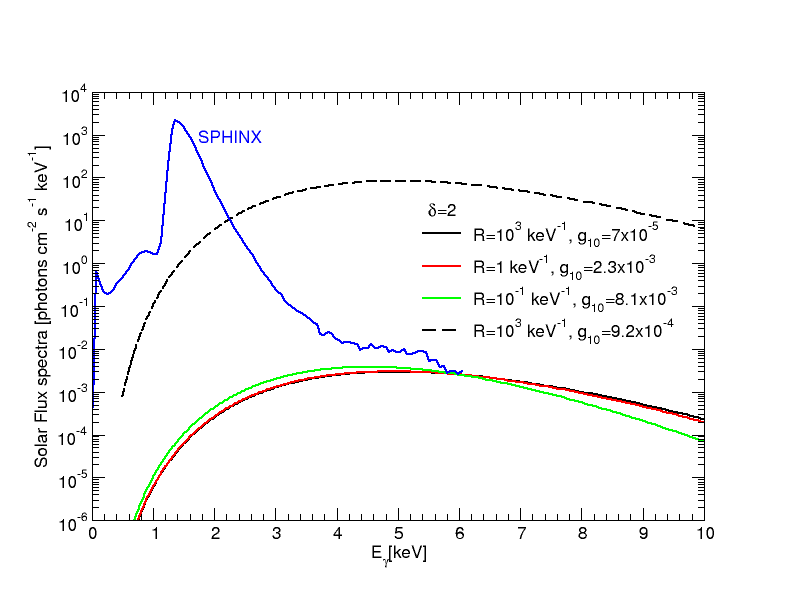}
%	\end{minipage}	
\end{tabular}
	\caption{The spectral photon flux from trapped KK axions as observed from Earth and the flux measured by SphinX. The top panel shows the case $\delta=1$ whereas $\delta=2$ is presented on the bottom one. Model parameters are indicated in the plots and they have been adjusted to remain below the SphinX measurements. The black dashed line on the bottom panel represents the photon flux obtained using the parameters from \cite{DiLella2003}.}
	\label{fig:luminosity}
\end{figure}

The spectral photon flux in a SphinX-like detector from the decay of trapped KK axions is presented in Figure \ref{fig:luminosity}, for $\delta=1,\,2$ and different values of the extra dimension radius\footnote{Notice that strictly speaking the curves for $R= 10^{-1}\, \kev^{-1}$ should start at $E=5$ keV, given that the lightest state in the KK tower would have a mass $m_1=10$ keV, but we include the full curves for comparison.}. Remaining below the SphinX measurements implies stringent limits on the maximal value of the axion-photon coupling. The set of parameters proposed by DiLella and Zioutas \cite{DiLella2003} is now ruled out. 

Measurements of the X-ray spectrum of the quiet Sun in Figure \ref{fig:luminosity} does not follow an isothermal distribution as expected from the element abundances. Significant deviations from the theoretical spectrum obtained with the CHIANTI database are observed for energies above $2.5\,\mathrm{keV}$ \cite{Sylwester2019}. Similar deviations are observed by RHESSI \cite{Hannah2010}. This could indicate the presence of non-thermal components (for instance via particle acceleration in solar microflares as measured by NuSTAR \cite{Glesener2020}), but it could also be explained by trapped KK axions as initially proposed by Zioutas \textit{et al.} \cite{Zioutas2004}. In Figure \ref{fig:luminosity}, one can see that the predictions of the photon flux from KK axions decay follow similar tendencies than the spectrum measured by SphinX, in particular for the case $\delta=1$. Measurements of the solar X-rays spectrum do not only constraints the KK axions model, they contribute to motivate it as a promising explanation for the non-thermal behavior of the solar X-rays emission.

Moreover, the decay of trapped KK axions would irradiate the Sun and was proposed \cite{DiLella2003} as an explanation for the heating of the solar corona. This hypothesis is supported by the abrupt temperature changes in the solar corona that share similar tendencies with the temperature fast increase in the Earth atmosphere under X-rays irradiation from the Sun \cite{Lean1997}. Such an explanation of the 80-year old solar corona problem remains plausible and is consistent with simulations of stellar atmosphere irradiation from a hotter X-rays source \cite{Madej2004}.

\subsection{Solar energy loss}
    
KK axions production in the solar core would result in energy loss and would consequently modify the helioseismological sound-speed profiles and the fluxes of neutrinos emitted by the Sun. Helioseismology has implied that the solar axion production would be indistinguishable from standard solar models as long as $L_a^D < 0.2\,L_\odot$ \cite{Schlattl1998}, value that has been used to establish the solar KK axion model \cite{DiLella2000, DiLella2003}. Solar neutrino fluxes have restrained to $L_a^D < 0.1\,L_\odot$ \cite{Gondolo2008} and a study combining helioseismology and neutrino emissions has set the limit $L_a^D < 0.03\,L_\odot$ \cite{Vinyoles2015}. The solar axion luminosity referred here is the direct luminosity from axions produced inside the Sun, which we distinguish from the trapped axion luminosity thanks to the superscript $D$ ($L_a^D$). 

The direct solar KK axion luminosity have been derived in \cite{DiLella2000},
\begin{equation}
	L_a^D = A\,L_\odot\,g_{10}^2\,\bigg(\frac{R}{\mathrm{keV}^{-1}}\bigg)^{\delta} \,,
\end{equation}
with $A=0.018$ for $\delta=1$, $A=0.19$ for $\delta=2$, and $A=2.1$ for $\delta=3$. We remind that $g_{10}=g_{\agg}\times 10^{10}\,\mathrm{GeV^{-1}}$. Using the limit $L_a^D < 0.03\,L_\odot$ we obtain the constraint from solar energy loss:
\begin{equation}
	g_{10}\,\bigg(\frac{R}{\mathrm{keV}^{-1}}\bigg)^{\delta/2} ~<~
\begin{cases}
1.3&\mathrm{for~\delta = 1}\,,\\
0.40&\mathrm{for~\delta = 2}\,,\\
0.12&\mathrm{for~\delta = 3}\,.\\
\end{cases}
\end{equation}
   
     \subsection{Extragalactic background light}
    
    The last constraint we will revise relies on the Extragalactic Background Light (EBL). Axions would be produced in the early Universe and in astrophysical objects, resulting in a large number of axions all around the Universe. The decay of an axion into two photons would then contribute to the luminosity of the Universe. In particular, measurements of the EBL put stringent limits on Axion-Like Particles (ALP) in the keV-range. One can compute the axion contribution to the EBL by matching the axion density with present observed DM density and taking into account the expansion of the Universe; then, if an axion decays, one must consider the redshift of the photons and the eventual absorption on atoms of the intergalactic medium. In this work, we extend the procedure of \cite{Cadamuro2012} to the case of KK axions. We consider the photoionization process on hydrogen and we neglect other processes as the Compton scattering or the pair production on atoms which do not play a role in the energy range considered \cite{Zdziarski1989}.
    
We need to compare the EBL measurements with the intensity of the photons flux from the KK axions decay denoted by $I_E$, multiplied by the energy to recover usual units. It is given by:
\be
E I_E = \int_{2 E_0}^\infty d m \frac{2 \pi^{\delta/2}}{\Gamma[\delta/2]} R^\delta m^{(\delta-1)} \frac{3 E_0 \rho(z_0,m)}{4 \pi m} x e^{-x} e^{-\kappa_{P}(E_0)} \label{eq:intensity} \,,
\ee
where the subindex ``0'' denotes values today, $x= \Gamma_{a\gamma\gamma} t_0(2E_0/m)^{3/2}$, $\rho(z_0,m)$ is the energy density today, and $\kappa_{P}(E_0)$ is the absorption factor due to photoionization process:
\bea
\kappa_{P}(z,E) &=& \int_0^{z} \frac{d z^\prime}{H(z^\prime) (1 + z^\prime)} n_H(z^\prime) \sigma_{P}(E)  \,,\label{absorp} \\
\sigma_{P}&=& 4.57\times 10^{-20} \left(\frac{10\,\kev}{E}\right)^{7/2}\,{\rm cm}^2 \label{eq:kappa} \,,
\eea
where $n_H$ is the hydrogen number density, $\sigma_{P}$ the hydrogen photoionization cross-section, $H$ the Hubble parameter and $z$ the redshift. In Eq.(\ref{eq:kappa}), we approximate the photoionization cross-section from \cite{Hubbell1969} by a power-law.

The main difference with standard ALPs lies in the superposition of states in the KK tower. To determine the EBL contribution due to KK axions, one must consider how the energy is distributed among the KK modes, which is given by the spectral function $\rho(z_0,m)$. This distribution depends on the production mechanism but also on mode-mixing: when the system starts oscillating, the energy gets redistributed among the whole KK tower. The system starts oscillating during the phase transition at which the axion gets a mass and, following the approach of \cite{Dienes2017}, one can parametrize the time dependent mass like:
\be
m(t)= \frac{m}{2} \left( 1 + {\rm erf}\big(\frac{1}{\sqrt{2 \delta_G}} \ln \frac{t}{t_G}\big) \right) \,, 
\ee
where $m$ is the late time axion mass, $m(t_G)= m/2$, and $\delta_G$  controls how fast  the transition takes place. In general, increasing $\delta_G$ suppresses the energy density in the KK tower; while increasing $m t_G$ leads to opposite effect, the energy density tends to be more democratically distributed. The authors of \cite{Dienes2017} define the parameter:
\be
\eta_{KK} = 1- {\rm max}_\lambda \frac{\rho_\lambda}{\rho}  \,,
\ee
where the subindex $\lambda$ refers to the mass eigenstate, and $\rho$ is the total energy density. Small values of $\eta_{KK}$ indicate that most of the energy density is stored practically in one mode.

The next step is to determine the mass distribution function $\rho(z_0,m)$. The derivation of this function should follow from integrating the EoM of the coupled systems, but such a detailed study is beyond the scope of this work. Instead, 
we proceed by making a plausible ansatz for this function by considering a Gaussian of width $\Delta_K$:
\bea
\rho(z_0,m)&=& \bar \rho_{KK}  f_G(\bar m, \Delta_K) \,, \\
f_G(\bar m, \Delta_K) &=& \frac{2}{\sqrt{\pi \Delta_K}} exp(-(\bar m)^2/\Delta_K)  \label{eq:fG} \,,
\eea
where $\bar m=m/m_1$, $m_1=1/R$, and $\bar \rho_{KK}$ is a parameter with units of energy density. We choose to normalize the function $f_G (\bar m, \Delta_K)$ to have unit area, such that:
\be
\rho_{KK} ~=~ \bar \rho_{KK} \frac{2 \pi^{\delta/2}}{\Gamma[\delta/2]} \int_{1}^{\infty} d \bar m {\bar m}^{(\delta-1)} f_G(\bar m, \Delta_K) ~\equiv ~  \bar \rho_{KK} I_\delta^{\infty} \,. 
\ee
We still need to relate the value of $\bar \rho_{KK}$ with the DM energy density today. Note that within the KK tower most of the modes will have lifetimes smaller than the age of the universe and will not contribute to the present DM energy density. The heaviest stable mode is given by:
\be
m_{DM} = \left( \frac{1.327\times 10^{-4}}{\gagg^2 \tau_u}\right)^{1/3} = \bigg(1.46\times 10^{-2} \,\kev \bigg)\,g_{10}^{-2/3}  \,,
\ee
and the fraction of KK modes that contribute to the DM energy density is:
\be
\rho_{KK}^{DM} ~=~ \bar \rho_{KK} \frac{2 \pi^{\delta/2}}{\Gamma[\delta/2]} \int_{1}^{\bar m_{DM}}  d \bar m {\bar m}^{(\delta-1)} f_G(\bar m, \Delta_K) ~\equiv ~  \bar \rho_{KK} I_\delta^{\bar m_{DM}} \,. 
\ee
Finally, using the present value of the DM energy density, we obtain: 
\bea
E I_E &=& \bigg(1.28916\times 10^9 {\rm nW}~ {\rm m}^{-2}~ {\rm str}^{-1}\bigg) \nonumber\\
& \times & \frac{2 \pi^{\delta/2}}{\Gamma[\delta/2]} \, g^2_{10}\, \frac{\eta_{KK}}{(1-\eta_{KK})I_\delta^\infty + \eta_{KK} I_\delta^{\bar m_{DM}}}\, \left(\frac{E_0}{m_1}\right)^{5/2} \left(\frac{m_1}{\kev}\right)^3\, F_{A\delta}\bigg\lbrace\frac{E_0}{m_1}, g_{10},m_1\bigg\rbrace \,, \nonumber\\
F_{A\delta}\bigg\lbrace\frac{E_0}{m_1}, g_{10},m_1\bigg\rbrace &=& \int_{\bar m_{min}}^\infty
d \bar m f_G(\bar m, \Delta_K) \bar m^{(\delta -1/2)} e^{-C_x \bar m^{3/2}} \, e^{-\kappa_{P}(E_0)} \,, \\
C_x &\equiv & 92\times g_{10}^2 \left(\frac{E_0}{m_1}\right)^{3/2} \left(\frac{m_1}{\kev}\right)^3 \,. 
\eea
Each KK mode cannot contribute to the spectrum with energies $E_0$ larger than half their mass, and therefore  $\bar m_{min}= 2 E_0/m_1$ when $2 E_0 \geq m_1$ and one otherwise. 

We are now ready to compare the expected light from KK axions decay to the EBL measurements. We approximately parametrize the upper limit of EBL published in \cite{Overduin2004} and \cite{Cooray2016} by the formula:
\be
E I_E^{max} \simeq 0.20412\times \lambda[\mathring{A}]^{0.47317}\, {\rm nW~m}^{-2}~{\rm str}^{-1} \label{eq:EBLmax} \,. 
\ee

\begin{figure}[t]
  \centering
  \begin{tabular}{c}
    \includegraphics[width=0.8\linewidth]{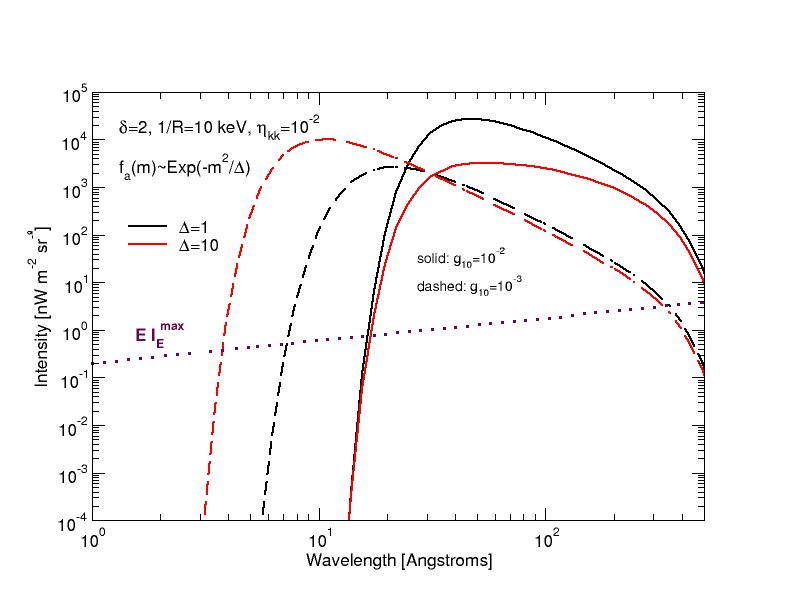} \\
    \includegraphics[width=0.8\linewidth]{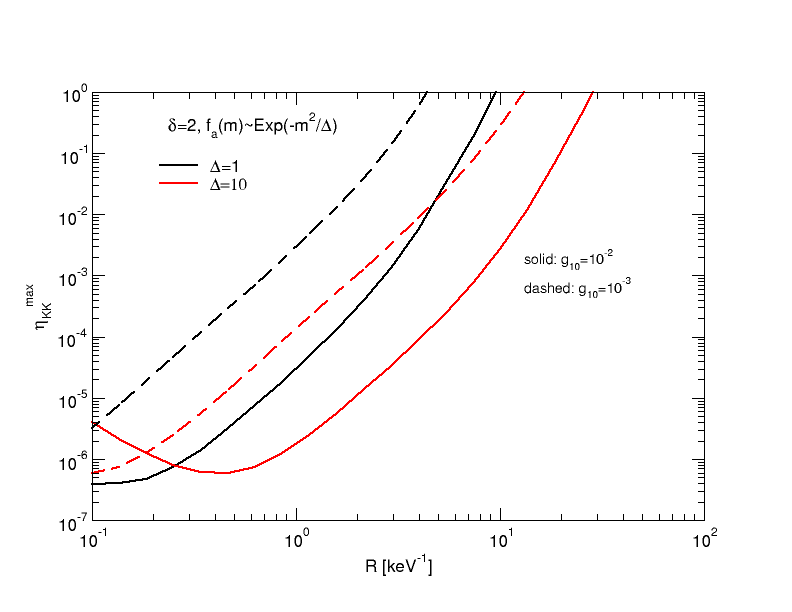}
    \end{tabular}
	\caption{Top panel: spectral intensity function from the decay of the KK tower for a set of parameters indicated in the plot. Solid lines correspond to $g_{a\gamma\gamma} = 10^{-12}\,\mathrm{GeV^{-1}}$ and dashed lines are obtained for $g_{a\gamma\gamma} = 10^{-13}\,\mathrm{GeV^{-1}}$. The purple dotted line $EI_E^{\rm {max}}$ corresponds to the experimental limit given by Eq.(\ref{eq:EBLmax}). Bottom panel: maximum allowed value of $\eta_{KK}$ in order to have $E I_E \leq E I_E^{max}$.}
	\label{fig:EBL}
\end{figure}

In Figure \ref{fig:EBL} are presented the EBL predictions from our model for several sets of parameters (top panel) and $\delta=2$ as an example. On the bottom panel we have plotted the maximum allowed value for $\eta_{KK}$ to guaranty that the contribution remains below the observational constraint as a function of the compactification radius $R$. In general, the smaller the width of the Gaussian, the easier to fulfil the EBL constraint since the energy density left in the tower diminishes. However, this trend is reverse when increasing $g_{\agg}$ (and therefore their decay rate into photons) and  $R \lesssim 1\, \kev^{-1}$, as shown on the bottom panel in Figure \ref{fig:EBL}. Nevertheless, while the EBL puts stringent limits in the case of standard ALPs in the energy range considered \cite{Cadamuro2012}, KK axions can easily escape this constraint, depending on how the energy is distributed among the KK tower. 
    
The revision and the update of the constraints that apply on the solar KK axion model, as detailed in this section, brings us to the determination of possible sets of parameters. Some of them are proposed in Table \ref{tab:parameters}. For instance, we see that the case $\delta=1$, $R=10^3\,\mathrm{keV}$, and $g_{10} = 9.8\times 10^{-4}$ fulfils the constraints and enables to have the fundamental gravitational scale, $M_\ast$, as well as the overall scale of the PQ symmetry breaking, $\bar{f}_{PQ}$, near the electroweak scale.

\renewcommand{\arraystretch}{1.3}
\begin{table}[H]
\centering
\begin{tabular}{ccccc||ccc}
\hline\hline
$(\delta\,,\, n)$ & $R~[\rm{keV^{-1}}]$ & $g_{10}$            & $\eta_{KK}$ & $\Delta_K$ & $M_\ast~[\mathrm{TeV}]$ & $f_{PQ}~[\mathrm{GeV}]$ & $\bar{f}_{PQ}~[\mathrm{GeV}]$ \\ \hline

$(1\,,\,2)$        & $5\times 10^{-1}$          & $7.3\times 10^{-3}$ & $8\times 10^{-6}$           & 1          & $2.0\times 10^3$       & $1.6\times 10^9$       & $6.4\times 10^2$             \\

$(1\,,\,2)$       & $10^{3}$          & $9.8\times 10^{-4}$ & --           & --          & $4.4\times 10^1$       & $1.2\times 10^{10}$       & $7.1\times 10^2$             \\ 

$(2\,,\,2)$       & $5\times 10^{-1}$          & $3.2\times 10^{-3}$ & $3\times 10^{-5}$          & 1          & $2.0\times 10^3$       & $3.6\times 10^{9}$       & $5.9\times 10^{-4}$             \\ 

$(2\,,\,3)$       & $5\times 10^{-1}$          & $3.2\times 10^{-3}$ & $3\times 10^{-5}$           & 1          & $5.4$       & $3.6\times 10^{9}$       & $2.1\times 10^{-1}$             \\ 

$(3\,,\,3)$        & $5\times 10^{-1}$          & $1.7\times 10^{-3}$ & $10^{-4}$           & 1          & $5.4$       & $6.8\times 10^{9}$       & $3.1\times 10^{-6}$             \\ 
\hline\hline
\end{tabular}
\caption{Possible sets of parameters allowed experimentally and the corresponding quantities defined in Section \ref{sec:axionLED}.}
\label{tab:parameters}
\end{table}

   %%%%%%%%%%%%%%%%%%%%%%%%%%%%%%%%%%%%%%%%%%%%%%%%%%%%%%%%%%%%%%

\section{Experimental searches for KK axion}\label{sec:searches}

Searching for standard axion represents a considerable challenge due to its feeble coupling to the particles of the Standard Model. As already stressed out, the situation changes drastically when considering large extra dimensions in which the axion can propagate. In this framework, the large multiplicity of the KK modes and their masses in the keV-range open a window for original detection strategies. This new point of view on the phenomenology could lead to a shift in the way of searching for axion, and the KK axion could be used as a probe for extra dimensions.

\subsection{Event rate of the KK axion decay in a  detector on Earth}
    
The first approach proposed to search for KK axions was the decay channel, $a\rightarrow\gamma\gamma$, since the decay could happen in a detector on Earth, the event rate depending not on the type of detector but only on its volume. The signature of the decay of a KK axion is the emission of two photons back-to-back, each of them carrying the same energy $E_\gamma = m/2$ since the gravitationally trapped axions are non-relativistic. In a Time Projection Chamber (TPC) with a 3D track reconstruction, as the MIMAC detector \cite{Santos2013} for instance, such a signature is almost unique, so it would lead to a very efficient background discrimination. Searching for KK axions via the detection of the photons emitted by the decay has been proposed in \cite{DiLella2000, Naples2004, Morgan2005, Battesti2007} and was used to set experimental constraints on solar KK axions \cite{XMASS, PacoThesis}. 
    
\begin{figure}[H]
	\centering
	\includegraphics[width=0.8\linewidth]{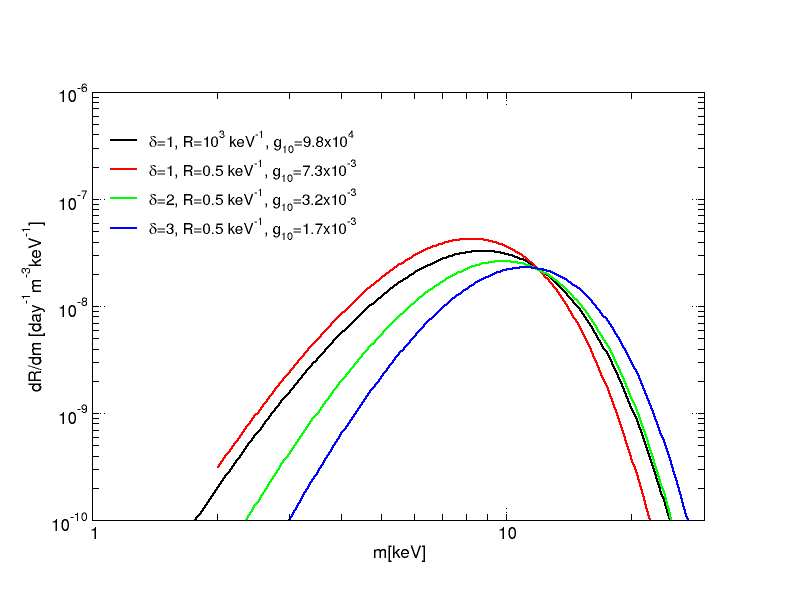}
	\caption{Differential decay rate of trapped KK axions into photons in a detector on Earth for the sets of parameters given in Table \ref{tab:parameters}.}
	\label{fig:decayRate}
\end{figure}

The differential decay rate of trapped KK axions into photons is given by:
\begin{equation}
    \frac{dR}{dm} ~=~ \Gamma_{a\gamma\gamma}\,\frac{dn_{KK}^{(T)}}{dm} ~=~ \Big(6.51\times 10^{-12}\,\mathrm{day^{-1}}\Big)\,g_{10}^2\,\bigg(\frac{m}{\mathrm{keV}}\bigg)^3\,\frac{dn_{KK}^{(T)}}{dm} \label{dRdm} \,,
\end{equation}
and is represented in Figure \ref{fig:decayRate}. The KK axion mass ranges from roughly $(1-30)\,\rm{keV}$ with an expected peak at $m = 2E_\gamma \simeq O(10)$ for each set of parameters. 

Integrating Eq.\eqref{dRdm} over the mass range we finally obtain the even rate. For the set of parameters in Table \ref{tab:parameters} and Figure \ref{fig:decayRate}, the event rate in a detector on Earth gives  $R \simeq 3\times 10^{-7}\,\mathrm{day^{-1}\cdot m^{-3}}$. The revision of the number density and the updated constraints from solar measurements have strongly reduced the event rate of $R \simeq 0.21\,\mathrm{day^{-1}\cdot m^{-3}}$ that was obtained from the DiLella and Zioutas model \cite{Morgan2005}.

This event rate forces to have a huge TPC-active volume requiring many observables to cope with the huge volume background. The high rejection power needed could be foreseen built on an upgraded MIMAC-type micro-TPC with 3D track reconstruction at few keV energies \cite{QuentinBkg, Tao2019}.

A last comment about other axion production channels. So far we have restricted our study to hadronic axion models, even though considering a non-zero axion-electron coupling, $g_{ae}$, offers additional mechanisms to produce KK axions in the Sun at a lower mass. The solar photon flux due to the electron-axion coupling will be proportional to $\gagg^2 g^2_{ae}$, while that due to coalescence is proportional to $\gagg^4$. And the same coupling dependence holds for the detection rate. So let us call $F_{\rm{SphinX}}$ the measured solar photon flux, and $F^{\gamma}= \gagg^4 \bar F^{\gamma}$ and $F^{e}=\gagg^2 g^2_{ae} \bar F^e$ the flux due to coalescence and axion-electron production respectively, such that:
\be
\gagg^2 ( \gagg^2 \bar F^{\gamma} + g^2_{ae} \bar F^e ) \leq F_{\rm{SphinX}} \,.
\ee
Similarly denoting by $R_{\gamma,\, e}$ the event rate from axion-photon or axion-electron coupling respectively, and using the "bar" notation to normalize quantities from values at unit coupling, one can then show that:
\be
  R= \gagg^2( \gagg^2 \bar R_\gamma  + g^2_{ae} \bar R_e )= \gagg^4\bar R_\gamma + (\frac{F_{\rm{SphinX}}}{\bar F^e}- \gagg^4 \frac{\bar F^\gamma}{\bar F^e}) \bar R_e
  \label{eq:compton} \,.
  \ee
Eq.(\ref{eq:compton}) tells us that considering the axion-electron coupling to produce KK axions via other processes in the Sun would only have minor influence on the constraint on $g_{a\gamma\gamma}$ imposed by measurements of the solar X-rays spectrum. In other words, any detector searching for KK axions via the coupling to photons will be limited by the constraint on $g_{a\gamma\gamma}$ from solar luminosity regardless of the production mechanisms.

    \subsection{Other strategies of detection}
    
   A promising approach to search for axions lies on the use of a magnetic field to convert an axion into a photon via the reverse Primakoff process. Such a detection strategy can be adapted to the case of KK axions. The CERN Axion Solar Telescope (CAST) has already set limits on KK axions in the frame of two additional dimensions \cite{Horvat2004, Lakic2008}. Since axion helioscopes are improving their sensitivities, dedicated searches for KK axions via the reverse Primakoff process could give interesting results. In a similar way, the experiments of "light shining through a wall" use the Primakoff effect in the aim of detecting the photon regeneration process $\gamma\rightarrow a \rightarrow \gamma$. Adapting the design of such experiments to the framework of large extra dimensions could open a detection channel for KK axions. 
   
   The search for KK axions can also be performed via indirect processes. In Section \ref{sec:constraints}, we have shown examples of how the phenomenology of the Sun or of the Extragalactic Background Light could be affected by the possible existence of KK axions. Burst of axions are expected in supernovae leading to a photon flux after axion decay or after axion-photon conversion. Oscillations between axions and photons are also expected, which would modify the photon fluxes from distant sources. Revisiting X-rays and $\gamma$-rays measurements in the framework of KK axions instead of standard ALP could lead to interesting results. We have shown that the trapped KK axions density evolves as $1/r^4$, with $r$ the distance to the Sun. Measurements of the solar photon flux at multiple distances to the Sun could reveal this $1/r^4$ behavior. For instance, the STIX telescope \cite{STIX} placed on Solar Orbiter will soon measure hard X-rays at distances to the Sun down to $0.22\,A.U.$.
   
   Finally, we have restrained our work to hadronic axion models in which the axions do not couple to fermions at tree level. Our study could be later extended by considering the axion-electron coupling, $g_{ae}$, that offers additional production mechanisms in the Sun \cite{Redondo2013} as well as other channels for detection. The main projects of dark matter direct detection have started to search for standard ALP since, for a non-zero axion-electron coupling, axions could induce events of the electron-recoils class \cite{LUX, Xenon}. Next generation of axion helioscopes, as the IAXO \cite{IAXO} for instance, could also improve their sensibility if axions couple to electrons. Once again, the existence of large extra dimensions would alter the phenomenology of the axions, and detection of KK axions via the axion-electron coupling appears as conceivable. 
    
\section{Conclusion}

The scenario of axions propagating in large extra dimensions addresses several theoretical issues $-$ as the hierarchy problem, the strong CP problem, or the nature of dark matter $-$ while offering a new scenario for explaining astrophysical observations. In this work we have entirely revised the solar KK axion model in which the particles are trapped in the gravitational field of the Sun. We here corroborate some results of the paper that pioneered the model \cite{DiLella2003}: (1) the number density of trapped KK axions evolves as $1/r^4$; (2) photon coalescence is the dominant production mechanism of trapped KK axions in the Sun; (3) the decay event rate $a\rightarrow\gamma\gamma$ in a detector on Earth is mainly constrained from measurements of the solar luminosity.

%The revision proposed in this paper meets some needs for experimental searches of KK axions since we provide analytical expressions of phenomenologically interesting quantities and we update the constraints with recent measurements.
The revision proposed in this paper aimed to provide analytical expressions of phenomenologically interesting quantities relevant for experimental searches of KK axions, and mainly to update the constraints with recent measurements.
We have computed the fluxes of KK axions produced in the Sun via the dominant processes in the case of hadronic models in which axions do not couple to fermions at tree level. The derivation of the number density of trapped KK axions represents the central element of the model. We obtain a density lowered by $\sim\mathcal{O}(50)$ compared with the previous estimation and we have cross-checked our result by an independent MC code provided on a public repository.

The properties of ALP in the keV-range are usually constrained by the measurement of the Extragalactic Background Light. However, depending on how the energy is distributed among the KK tower, the KK axions can escape this cosmological bound. The most stringent constraint on the model comes from the X-rays measurements of the quiet Sun. This limit on the decay event rate in a detector cannot be avoided: the amount of KK axions decaying on Earth is correlated to the amount of KK axions that decay in between the Earth and the Sun and which contribute to the observed solar luminosity. We show that any strategy of detection making use of the axion-photon coupling, $g_{a\gamma\gamma}$, will experience this astrophysical bound, even when considering production mechanisms via the axion-electron coupling. We derive a decay event rate of about $3\times 10^{-7}\cdot\mathrm{day^{-1}\cdot m^{-3}}$ which lies $6$ orders of magnitude below previous estimations \cite{DiLella2003,Morgan2005}, with major consequences on current experiments searching for KK axions via this channel. 

The X-rays measurements of the quiet Sun strongly limit the decay event rate in a detector though they also bring additional motivation for searching for KK axions. We have shown that the non-thermal component of the X-rays distribution could possibly be explained by the decay of trapped KK axions. This appealing hypothesis should be further investigated, for instance by studying the physics of KK axions during solar microflares or by upgrading our analysis with upcoming STIX measurements at $0.22\,\mathrm{A.U.}$. Further work could also extend the study of KK axions to non-hadronic models that offer additional production mechanisms in the Sun \cite{Redondo2013} and open other channels for detection. The axion-electron coupling could play an interesting role in solar microflares.

Searches for KK axions via the decay in a detector on Earth appears as very challenging for present TPCs. However, complementary strategies of detection, from direct or indirect approaches, could be used in the near future to search for KK axions. The framework of large extra dimensions offers new possibilities for the physics of the axion. Current experiments searching for ALP could be revisited and extended to this scenario. The axion could then be used as a probe for extra dimensions.

\acknowledgments{We wish to thank Barbara and Janusz Sylwester who kindly shared with us the SphinX measurements of solar X-rays and who answered with details our questions about the working principles of the SphinX detector. MBG would like to thank the LPSC (UGA, CNRS, Grenoble INP) for kind hospitality. The work of MBG has been partially supported by MICINN (PID2019-105943GB-I00/AEI/10.13039/501100011033) and ``Junta de Andaluc\'ia" grants  P18-FR-4314  and  A-FQM-211-UGR18.
}

%%%%%% APPENDIX
\appendix

\section{Plasmon decay rate}
\label{app:plasmonDecay}

To compute the rate of KK axion production from the plasmon decay $\gamma_T\rightarrow a + \gamma_L$ we need the thermal decay rate:
\be
\Gamma_{\gamma_T\rightarrow a\gamma_L} (T) = \frac{1}{2E} \int \frac{d^3 k_T}{(2 \pi)^3 2 \omega_T}  \frac{d^3 k_L}{(2 \pi)^3 2 \omega_L} (2 \pi)^4 \delta^4(k_T - k_L - p) \big|M_{\gamma_T\rightarrow a\gamma_L}\big|^2 (1+ f^{eq}(\omega_T)) \label{thermaldecTL}\,,
\ee
where \cite{RaffeltPlasmon}:
\be
\big|M_{\gamma_T\rightarrow a\gamma_L}\big|^2=g_{\agg}^2 \frac{\Big|(\vec e_T \times \vec k_T)\cdot \vec k_L\Big|^2}{k_L^2} \frac{\omega_P^2}{k_L^2 + \omega_P^2} \omega_L^2 =
g_{\agg}^2 \frac{|\vec k_T \times \vec k_L|^2}{2k_L^2} \frac{\omega_P^2}{k_L^2 + \omega_P^2} \omega_L^2 
\,,
\ee
and $\delta^4 ( k_T- k_L -p) = \delta^3 (\vec{k}_T-\vec{k}_L- \vec{p}) \delta ( \omega_T - \omega_L -E)$ enforced energy and momentum conservation:
\bea
E&=& \omega_T- \omega_L \,,\\
\vec{p}&=& \vec{k}_T - \vec{k}_L\,, \\
k_L^2 &=& k_T^2 + p^2 - 2 k_T p \cos \theta_T \,. 
\eea
 Neglecting the thermal correction for the time being (\textit{i.e.}, taking $f^{eq}(\omega_T) \ll 1$), we have:
\bea
2 E \Gamma_{\gamma_T\rightarrow a\gamma_L} (T) &=& \frac{1}{2 \pi^2} \cdot \frac{1}{4 \omega_L \omega_T}\int d^3 k_T  \delta(\omega_T - \omega_L - E)\, \big|M_{\gamma_T\rightarrow a\gamma_L}\big|^2 \nonumber \\
&=& \frac{g^2_{\agg}}{2 \pi^2} \cdot \frac{\omega_L^2}{4 \omega_L \omega_T}\int d\Omega_T \frac{k^2_T}{|d W/dk_T|}\cdot \frac{\omega_P^2}{k_L^2 + \omega_P^2}\cdot
\frac{k_T^2 p^2 ( 1 - \cos^2 \theta_T)}{k_L^2} \nonumber \\
&=& \frac{g^2_{\agg}}{32 \pi} \cdot \frac{\omega_L^2 \omega_P^2 k_T^2}{\omega_T p}\int d\cos \theta_T  \frac{1 - \cos \theta_T^2}{(A - \cos \theta_T)(B - \cos \theta_T) (C - \cos \theta_T)} \,, \label{EGammaTL}
\eea
where in the second line we have: 
\bea
W&=&\omega_T - \omega_L - E \,, \\
\frac{dW}{d k_T}&=& \frac{k_T}{\omega_T}- \frac{k_L}{\omega_L}\frac{d k_L}{d k_T}
%\frac{(\omega_L + \omega_T) k_T}{\omega_L \omega_T} \left( 1 - \frac{\omega_T}{\omega_T + \omega_L} \frac{p}{k_T} \cos \theta_T \right) \nonumber \\
= \frac{p}{\omega_L} \left( \frac{E}{\omega_T} \cdot \frac{k_T}{p} - \cos \theta_T \right) \,, 
\eea
and in the last line in \eqref{EGammaTL} we have defined:
\bea
A &=& \frac{E}{\omega_T} \cdot \frac{k_T}{p} \,, \\
B &=& \frac{1}{2} ( \frac{k_T}{p} + \frac{p}{k_T} ) \,, \\
C &=& B + \frac{\omega_P^2}{2 k_T p} \,.
\eea

For slow-moving axions, in the limit $p/k_T \ll 1$ we have:
\be
2 E \Gamma_{\gamma_T\rightarrow a\gamma_L} \simeq \frac{g^2_{\agg}}{16 \pi} \frac{\omega_P^2 \omega_L^2 p^2 }{k_T E} \cdot \frac{k_T^2}{k_T^2 + \omega_P^2} \int  \int d\cos \theta_T  (1 - \cos \theta_T^2)
\simeq \frac{g^2_{\agg}}{12 \pi} \frac{\omega_P^2 \omega_L^2 p^2 }{k_T E} \cdot \frac{k_T^2}{k_T^2 + \omega_P^2} 
\,,
\ee
and therefore in this limit:
\be
\frac{d N_a}{d E} \simeq \frac{f_a^{eq}(E)}{2 \pi^2} 
\frac{g^2_{\agg}}{24 \pi} \frac{\omega_P^4 k_T }{k_T^2 + \omega_P^2} \cdot \frac{p^3}{E} \,, \label{dnaTL}
\ee
where we have replaced $\omega_L^2 \simeq \omega_P^2$. 

%%%%%%%%%%%%%%%%%%%%%%%%%%%%%%%%%%%%%%
\section{Derivation of the number density}\label{app:density}

We start by deriving the gravitational potential inside the Sun from the Poisson equation:
 \bea
 \Phi_G(r < R_\odot) &=& -4 \pi G \left( \frac{1}{r}\int_0^{r} {r^\prime}^2 \rho(r^\prime) d r^\prime + \int_r^R r^\prime \rho(r^\prime) dr^\prime \right) \nonumber \\
 &=& -\frac{G M}{R_\odot} \left( 1 + \int_{\bar r}^1 \frac{\bar M}{\bar r^2} \right) d \bar r = - \frac{G M}{R_\odot} \bar \Phi_G(\bar r)\,,
 \eea
 where in the second line ``bar'' quantities are normalized by their Sun value. 
 Bounded trajectories follow from the condition $v_r=\dot r=0$, which can be written as:
 \be
 v_{r,0}^2 + r_0^2 {\dot \phi_0}^2 ( 1 -\bar r_0^2 \bar u_r^2) = \frac{2 GM}{R_\odot} ( \bar \Phi_G(\bar r_0) - \bar u_r)\,, \label{bounded}
 \ee
 where the subindex ``0'' refers to the initial condition for radius and velocities inside the Sun, and $\bar u_r=R_\odot/r$. In terms of the initial kinetic energy per unit mass, $\bar E_K= (v_{r,0}^2 + r_0^2 \dot \phi_0^2)/2$, and using $a^2= v_{r,0}^2/(2 \bar E_K)$, we have:
 \be
 \bar E_K (\bar r_0)= \frac{GM}{R_\odot} \cdot \frac{\bar \Phi_G(\bar r_0)- \bar u_r}{ 1 - \bar r_0^2 (1 -a^2) \bar u_r^2/2} \,.
 \ee
 Bounded orbits at a radial distance $r$ are obtained for particles produced inside the Sun with kinetic energy in the range $( E_K^{min}, E_K^{max})$ depending on the initial angular momentum. To reach  the Earth, for particles produced at the Sun surface, their kinetic energy must be in the range [0.995363, 0.995374] in units of $GM/R_\odot= 2 v_{esc}^2$; for those produced at half the Sun radius we have instead $E_K \in [1.952736, 1.95274]$.

 In order to get the number density of trapped KK axions, we start again with the Boltzmann equation for axions produced inside the Sun due to photon coalescence. We neglect the other production processes since they are suppressed for low momenta (\textit{c.f.} Section \ref{sec:axionProd}):
 \be
 \frac{dn_a}{dt}= \frac{1}{(2 \pi)^3}\int d^3 p \Gamma_{a\gamma} f_a^{eq}(E) \,, \label{dnadt0}
 \ee
 where for the rate of production we use the simplification of Eq.(\ref{eq:coalRateSimplify}):
 \be
\Gamma_{a \gamma} \simeq \Gamma_{\agg} \frac{m}{E} \,.
\ee
 Restricting the momentum integral to those values which give bounded orbits at a distance $r$ we obtain the number density of trapped KK axions. They will have small velocities and behave as non-relativistic particles, and therefore:
\be
d^3p = d p_x dp_y dp_z \simeq m^3 v^2 dv d\phi d\cos \theta \,.
\ee
Due to isotropy, we can perform the integral along $\theta$ and work in the $XY$ plane:
\be
d^3 p= 2 m^3 v^2 dv d\phi = 2 m^3 \frac{v^2}{\sqrt{v^2 - v_\phi^2}} dv dv_\phi \,.
\ee

We can now perform the integral in velocities with the constraint for bounded trajectories Eq.(\ref{bounded}), which gives the required velocity at production at a radius $r_0$ inside the Sun to get to a distance $r$ from the center of the Sun. Integrating over the solar volume we finally obtain the number density of trapped KK axions $n_a^{(T)}$ per unit time: 
\be
\frac{d n_a^{(T)}}{dt} = \frac{m^3}{(2 \pi)^3} 4 \pi R_\odot^3 \int_0^1 \bar r_0^2 d \bar r_0 \int_0^\infty dv \int_0^v d v_\phi \frac{4 v^2}{\sqrt{v^2 - v_\phi^2}} \cdot N_\delta\, \delta( f_T(v, v_\phi) ) \cdot \Gamma_{a\gamma}(T) f_a^{eq}(E) \,,
\label{dnaTdt}
\ee
where
\be
f_T(v, v_\phi)=
v_r^2 + v_\phi^2 ( 1 -\bar r_0^2 \bar u_r^2) -\frac{2 GM}{R_\odot} ( \bar \Phi_G(\bar r_0) - \bar u_r) \,.
\ee
The function $N_\delta \delta (f_T)$ is the probability density for a particle of energy $E(v,v_\phi)$ to be at radius r, such that
\be
\int d^3 r N_\delta \delta(f_T) =1 \,.
\ee
Let us consider for the time being only radial trajectories to simplify the analyses, \textit{i.e.}, $v_\phi=0$. The constraint Eq.(\ref{bounded}) simplifies and the maximum radial distance of the bounded trajectory is given by:
 \be
 \bar{r}_B= \frac{2}{ 2 \bar \Phi_G(\bar r_0) -\bar{v}^2 } \,,
 \ee
 where $\bar v^2= v^2/(GM/R_\odot)$, $\bar{r}_B=r_B/R_\odot$,  
 and the normalization constant of the probability density is:
 \be
 N_\delta = \frac{1}{32 \pi R_\odot^3} (2 \bar \Phi_G(\bar r_0) -\bar{v}^2)^4 \,.
 \ee
 We can now perform the integral over velocities to get the number of trapped axions:
\be
I_v(\bar r_0, \bar r)=  R_\odot^3 \left( \frac{G M}{R_\odot}\right)^{3/2} \int_{\bar v_{min}}^{\bar v_{max}} d \bar v \bar v^2 N_\delta \delta(f_T)= \frac{1}{2 \pi \bar r^4} \sqrt{B(\bar r,\bar r_0)}\,, \label{Ivdelta}
\ee
where we have defined:
\be
B(\bar r,\bar r_0)= 2 (\bar \Phi_G(r_0) -\bar u_r) \,.
\ee
The number density per unit time of trapped KKaxions is then:
\be
\frac{d n_a^{(T)}}{dt} = \frac{m^3}{4 \pi^3 \bar r^4} \int_0^1 \bar r_0^2 d \bar r_0 \sqrt{B(\bar r,\bar r_0)} \cdot \Gamma_{a\gamma}(T) \cdot f_a^{eq}(E) \,,
\label{dnaTdtPdel}
\ee
where $E \simeq  m \big(1 + \frac{GM}{2R_\odot} \sqrt{B(\bar r,\bar r_0)}\big)$. This gives the $1/r^4$ behavior as in \cite{DiLella2003}. 

Following \cite{VanTilburg}, the probability density is inversely proportional to the radial velocity, so instead of a delta function to implement the constraint of bounded trajectories, one must use:
\be
P(\bar r, \bar r_0) = \frac{N_P}{\sqrt{\bar v^2 - B(\bar r,\bar r_0)}} \,,
\ee
where again imposing:
\be
4 \pi R_\odot^3 \int d \bar r \bar r^2 P(\bar r, \bar r_0) = 1 \,,
\ee
one gets:
\be
R_\odot^3 N_P= \frac{1}{10 \pi^2} (2 \bar \Phi_G(\bar r_0) - \bar v^2)^{7/2} \,.
\ee
Because trapped axions are non-relativistic, one can approximate $E\simeq m$, and as before perform the velocity integral:
\be
I_v(\bar r_0, \bar r)=  \int_{\bar v_{min}}^{\bar v_{max}} d \bar v \bar v^2 \frac{1}{10 \pi^2} \frac{(2 \bar \Phi_G(\bar r_0) - \bar v^2)^{7/2}}{(\bar v^2 - B(\bar r, \bar r_0))^{1/2}} \,,  \label{IvPr}
\ee
which can be done numerically and again gives the $1/r^4$ behavior for the trapped axions.

The last step is to sum over the KK-tower, and integrate over time  with the decay term from the Boltzmann equation. For a KK tower of massive axions, the lifetime of some modes can be smaller than the Sun age, so one must include production and decay at the same time in the Boltzmann equation. Taking this into account and summing over the tower of KK axions, we obtain for the number density of trapped axions\footnote{To recover the number density of axions in cm$^{-3}$, we recall that  $GM/R_\odot$ is $GM/R_\odot/c^2=2.11119\times 10^{-6}$, and with all other quantities in units of energy, \textit{i.e.} keV, we get $n_a$ in units of $\kev^3$. Dividing by $(\hbar c)^3$ we convert to cm$^{-3}$.}:
\bea
n_{KK}^{(T)}&\simeq& \frac{1}{2 \pi^2} \cdot
  \left( \frac{G M}{R_\odot}\right)^{3/2} \cdot
  \frac{2 \pi^{\delta/2}}{\Gamma[\delta/2]} R^\delta \cdot \frac{\gagg^2}{64 \pi}
  \int dm \frac{m^{(\delta +5)}}{\Gamma_{\agg}} ( 1 - e^{-t_\odot \Gamma_{\agg}}) \nonumber \\ 
  & &\times \int_0^1 \bar r_0^2 d \bar r_0 \cdot f_a^{eq}(m) \frac{m}{E}I_v[\bar r, \bar r_0] \,, \label{naTKK}  
\eea
where the function $I_v$ is given either by Eq.(\ref{Ivdelta}) or Eq.(\ref{IvPr}) depending on the probability density considered.

Finally, collecting all factors:
\be
n_{KK}^{(T)}(t, r) \simeq 2.2 \times 10^{14} {\rm cm}^{-3} 
\cdot g_{10}^2 \cdot I[\bar r, \delta] \,,
\ee
where $g_{10}=g_{\agg}\times 10^{10}$ GeV$^{-1}$,  and the integral is given by\footnote{When doing the integral, $\Gamma_{\agg} t_\odot$ is dimensionless, i.e., if $t_\odot$ is given in $s$ then $\Gamma_{\agg}$ is given in $s^{-1}$. }:
\be
I[\bar r, \delta]= \frac{2 \pi^{\delta/2}}{\Gamma[\delta/2]} \int dm (R m)^\delta m^5 \cdot \frac{1 - e^{-t_\odot \Gamma_{\agg}}} {\Gamma_{\agg} t_\odot} 
\times \int_0^1 \bar r_0^2 d \bar r_0 \cdot f_a^{eq}(m) \frac{m}{E}\cdot I_v[\bar r, \bar r_0] \,,
\ee
where $m$ the axion mass, $R^{-1}$, $T$ and $E$ are given in keV.

%%%%%%%%%%%%%%%%%%%%%%%%%%%%%%%%%%%%%%%%%%%%%
\section{KK axions within the FoV of SphinX}\label{app:FoV}

Given the number density of trapped KK axions:
\be
n_a^T(r,m) \simeq \frac{n_a^T(R_\odot,m)}{\bar r^4} \,,
\ee
the number of trapped KK axions within the detector FoV is given by:
\bea
N_a^T(m)&=& \int_{FoV}d {\bf r}^3 \frac{r^2}{4 \pi D^2} n_a^T(r,m) =
\frac{2 \pi}{4 \pi}\int_0^{\infty} d ~D \int^{1}_{\cos \alpha_0} d \cos \alpha ~n_a^T(r,m) \nonumber \\ 
&\simeq &   \frac{R_\odot n_a^T(R_\odot,m)}{2 \bar r_D^3} \int_0^\infty dx \int_{y_0}^1 dy \frac{1}{(x^2 + 1 - 2 x y)^2}  \,, \label{NaEarthr0}
\eea
where we perform the integral from the ``Earth'' reference system with $y=\cos\alpha$, $x=\bar D/\bar r_D$,  and consider the Sun as a point source. However, we have to subtract the Sun contribution when $\bar r < 1$; \textit{i.e.}, we have to take the limits of integration such that:
\be
x^2 - 2 x y + 1 > \frac{1}{\bar r_D^2} \,.
\ee
The roots of this equation are given by:
\be
x_\pm(y) = y \pm \sqrt{y^2 - \Delta^2} \,,
\ee
where $\Delta^2= 1 - 1/\bar r_D^2$, $x_\pm(1)= 1\pm \sqrt{1-\Delta^2}$ and $x_\pm(\Delta)= \Delta$. Therefore, we have now the integral:
\be
I^E = \int_0^{\infty} d x \int_{y_0}^1 dy \frac{1}{(x^2 + 1 - 2 x y)^2} - \int_\Delta^1 dy \int_{x_-}^{x_+}dx \frac{1}{(x^2 + 1 - 2 x y)^2}= I^E_{All} - I^E_{Sun} \,.
\ee
Doing first the integral in ``$x$'', we have: 
\bea
I_{All}^E[y] &=& \int_0^\infty d x\frac{1}{(x^2 + 1 - 2 x y)^2} \nonumber \\
&=& \frac{1}{2} \left( \frac{\pi/2}{(1-y^2)^{3/2}} + \frac{y}{1-y^2}+ \frac{1}{(1-y^2)^{3/2}} \arctan{\frac{y}{\sqrt{1 -y^2}}}\right) \,,
\eea
and therefore:
\bea
I^E_{All} &=& \int_{y_0}^1 dy I_{All}^E[y] = \frac{1}{2} \left(
\frac{y}{\sqrt{1-y^2}}\cdot \frac{\pi}{2} + \frac{y}{\sqrt{1-y^2} }\cdot \arctan{\frac{y}{\sqrt{1-y^2}}} \right)_{y_0}^1 \nonumber \\
&=& \frac{1}{2} \left( \underset{{ \alpha \to 0}}{\rm limit} \frac{\pi}{\tan \alpha} - \frac{\pi}{\tan \alpha_0} -1 + \frac{\alpha_0}{\tan \alpha_0}
\right) \,.
\eea
Similarly, for the Sun contribution we have:
\bea
I_{Sun}^E[y] &=& \int_{r_+}^{r_-} d x\frac{1}{(x^2 + 1 - 2 x y)^2}
= \frac{\sqrt{y^2-\Delta^2}}{(1-\Delta^2)(1-y^2)}+ \frac{1}{(1-y^2)^{3/2}} \arctan{\frac{\sqrt{y^2-\Delta^2}}{\sqrt{1 -y^2}}} \\
I_{Sun}^E &=& \int_\Delta^1 d y I_{Sun}^E[y] =
\left(\frac{y}{\sqrt{1-y^2}} \arctan{\frac{\sqrt{y^2-\Delta^2}}{\sqrt{1 -y^2}}} - \frac{\Delta^2}{1-\Delta^2} \ln (y + \sqrt{y^2-\Delta^2})\right)_\Delta^1 \nonumber \\
&=& 
\underset{{ \alpha \to  0}}{\rm limit} \frac{\pi/2}{\tan \alpha} - \frac{1}{\sqrt{1-\Delta^2}} - \frac{\Delta^2}{1-\Delta^2} \ln \frac{1+\sqrt{1-\Delta^2}}{\Delta} \,.
\eea
The divergence at the Sun center (when $\alpha \rightarrow 0$) cancels out, 
and finally we get:
\bea
N_a^T(m) &=& \frac{R_\odot n_a^T(R_v,m)}{2 \bar r_D^3} \left( \bar r_D + (\bar r_D^2-1) \ln \frac{1 + \bar r_D}{\sqrt{\bar r_D^2-1}}- (1 - \frac{(\alpha_0-\pi)}{\tan \alpha_0}) \right) \nonumber \\
&\underset{\bar r_D \gg 1}{\simeq}& \frac{R_\odot n_a^T(R_\odot,m)}{2 \bar r_D^3} \left( 2\bar r_D -\frac{2}{3\bar r_D}- \frac{1}{2}(1 - \frac{\alpha_0-\pi}{\tan \alpha_0}) \right) \,.\nonumber \label{NaTEarth2} \\
\eea

%%%%% PRINT BIBLIO
\bibliographystyle{JHEP}
\bibliography{biblio}

\end{document}